\documentclass[12pt]{article}
\usepackage{verbatim,vmargin,graphicx,amsmath,calc}
\usepackage{multirow}
\usepackage{epsfig}
\usepackage{ifthen}
\usepackage{alltt}
\usepackage{rotating}
\usepackage[super]{nth}

\usepackage{color}
\usepackage{tikz}
\usepackage{listings}
\usepackage{hyperref}    % Hyperlinks in references
\usepackage[all]{hypcap} % Internal hyperlinks to floats.
\usepackage{cite}

\input{babarsym}

\newcommand{\KPP}         {\mbox{$\Kpm  \pimp \pipm$}}
\newcommand{\BtoKPP}      {\mbox{$\Bpm \to \KPP$}}
\newcommand{\mACSq}       {\mbox{$m^2_{K\pi}$}}
\newcommand{\mBCSq}       {\mbox{$m^2_{\pi\pi}$}}
\newcommand{\rhoz}        {\mbox{$\rho^0$}}
\newcommand{\rhoI}        {\mbox{$\rhoz(770)$}}
\newcommand{\dxrho}       {\mbox{$\Delta x_{\rho}$}}
\newcommand{\jfit}        {\mbox{\textbf{\textit{J}}\hspace{-0.15em}\textsc{fit}}\xspace}
\newcommand{\sjfit}       {\mbox{\textbf{\textit{\scriptsize J}}\hspace{-0.1em}{\tiny \rm FIT}}\xspace}
\newcommand{\dxrhogen}{\mbox{$\Delta x_{\rho}^\mathrm{Gen.}$}}
\newcommand{\dxrhoavg}{\mbox{$\Delta x_{\rho}^\mathrm{Avg.}$}}
\newcommand{\dxrhojfit}{\mbox{$\Delta x_{\rho}^{\sjfit}$}}
\def\root       {\mbox{\textsc{Root}}\xspace}
\def\roofit     {\mbox{\textsc{RooFit}}\xspace}
\def\laura      {\mbox{\textsc{Laura$^{++}$}}\xspace}
\def\cpp        {\mbox{\textsc{C\raisebox{0.1em}{{\footnotesize{++}}}}}\xspace}

\begin{document}

%\begin{flushleft}
%\today \\[1cm]
%\end{flushleft}

% Title of the paper
\begin{center}
{\Large 
\jfit{\bf : a framework to obtain combined experimental results through joint fits}
}
\end{center}
% end title
\smallskip

\begin{center}
{\large E. Ben-Haim$^1$, R. Brun$^2$, B. Echenard$^3$, T.E. Latham$^4$}
\vskip 0.5cm
{\it $^1$Laboratoire de Physique Nucl\'eaire et de Hautes Energies (LPNHE),
IN2P3/CNRS, Universit\'e Pierre et Marie Curie-Paris 6,
Universit\'e Denis Diderot-Paris 7} \\
{\it $^2$European Organization for Nuclear Research (CERN), Geneva, Switzerland}\\
{\it $^3$California Institute of Technology, Pasadena, California 91125, USA}\\
{\it $^4$Department of Physics, University of Warwick, Coventry CV4 7AL, United Kingdom}
\end{center}

\begin{abstract}

A master-worker architecture is presented for obtaining combined experimental results through joint fits of datasets from several experiments.
The design of the architecture allows such joint fits to be performed keeping the data separated, in its original format, and using independent fitting environments.
This allows the benefits of joint fits, such as ensuring that correlations are correctly taken into account and better determination of nuisance parameters, to be harnessed without the need to reformat data samples or to rewrite existing fitting code.
The \jfit framework is a \cpp implementation of this idea in the \laura package, using dedicated classes of the \root package.
We present the \jfit framework, give instructions for its use, and demonstrate its functionalities with concrete examples.

\end{abstract}

%\smallskip

\section{Introduction}
\label{sec:intro}

In the vast majority of cases, high energy physics experiments obtain measurements by performing a minimum chi-square or a maximum-likelihood fit to their data to extract observables of interest.
Combining results of different experiments is routinely performed, using different well-established methods and dedicated tools.
In the simplest approach, which is often used, measured observables are assumed to be normally distributed and standard statistical prescriptions are readily applied.
However, there are situations in which the procedure is challenging even with the simple normal-distribution hypothesis, in particular for measurements involving a large number of parameters (including nuisance parameters).
The size of the covariance matrices, when available, makes the procedure tedious and prone to errors.
When non-Gaussian uncertainties are taken into account the combination procedure becomes much more complicated, as working with the actual likelihood or chi-square functions is generally needed.
This is indispensable in many complex measurements, in cases with small sample sizes and combination of upper limits.
Often only a partial likelihood or chi-square function is considered, which is the projection of the full function on particular parameters of interest, assuming that they are uncorrelated with the other parameters.
Ideally, combining measurements should be done by fitting simultaneously the different data samples by means of a joint fit.

The straightforward way to perform joint fits involves collecting data in a common format, and has been implemented, among others, in the \roofit package~\cite{ref:roofit}.
This solution could however be inefficient if dedicated fitting frameworks have already been developed, especially for complex fit models.
It may also raise issues with data access policies of each experiment. 
An alternative approach consists of performing a joint fit while keeping the data separated.
This can be achieved using a master-worker architecture, in which the master drives the fit by combining the 
values of the likelihood functions returned by several workers, each of which is specific to an experiment and accesses its own dataset. 

In this paper we present \jfit: an implementation of this idea within the \laura package~\cite{ref:laura} that uses several classes from the \root framework~\cite{ref:root}.
In Sec.~\ref{sec:JointFitsAndExamples}, we revisit the general formalism of joint fitting, and briefly advocate its advantages.
In Sec.~\ref{sec:jfitFramework} we explain the concept of the master-worker architecture and describe in detail the \jfit implementation of such an architecture in the \cpp programming language.
We also provide details of how to minimally adapt existing fitting codes in order to use them within the framework.

\section{Joint fits and their benefits}
\label{sec:JointFitsAndExamples}

The formalism to combine several measurements by performing a joint maximum-likelihood fit\footnote{The formalism is discussed here in terms of likelihood, but the benefits of joint fitting also apply when using a minimum chi-square fit.} of 
different datasets (e.g., from different experiments) is well known:
the likelihood $\cal{L}$ to maximise is given by the product of likelihoods from the different datasets.\footnote{
A brief review of the maximum-likelihood estimation method is given in Appendix~\ref{max}.}
In the case of two datasets $A$ and $B$ (the generalisation to a larger number of datasets is straightforward), with individual likelihoods ${\cal L}_A$ and ${\cal L}_B$, the combined likelihood is written as: 
\begin{equation}
{\cal L}(\theta, \theta_A,\theta_B) = {\cal L}_A(\theta, \theta_A){\cal L}_B(\theta, \theta_B) = \left[ \prod_{x_{i\phantom{j}}\! \in \, x_A}{\cal P}_A(x_i; \theta, \theta_A) \right] \left[ \prod_{x_j \, \in \, x_B}{\cal P}_B(x_j; \theta, \theta_B) \right],
\label{eq:combL}
\end{equation}
where $x$ is a random variable (or a set of several random variables) with corresponding ensembles of independent observations, designated by $x_A$ and $x_B$, in the two datasets; ${\cal P}_A(x,\theta, \theta_A)$ and ${\cal P}_B(x,\theta, \theta_B)$ are the probability density functions followed by $x$ in the two datasets; $\theta$ denotes the parameters of interest that are shared by the two datasets, and have to be simultaneously extracted from both (common parameters); $\theta_A$ and $\theta_B$ are parameters that are specific for $A$ and $B$ (specific parameters).

Joint fitting to combine results from two experiments has many advantages compared to na\"ive averaging methods. 
These benefits mainly derive from the fact that joint fits take into account the correlations between all fit parameters, not just those of primary interest.
The exploration of the full likelihood surface, rather than some projection of it, can result in a more reliable estimation of both the central value and the statistical uncertainties.
These uncertainties can also be reduced because common nuisance parameters can be better constrained, which leads to improved precision on the parameters of interest.

The estimation of systematic uncertainties is also made more straightforward.
In particular, uncertainties originating from an external source, such as a measured property of a background process, can be accounted for simply by repeating the joint fits using a set of modified assumptions about that external input.
Furthermore, the need to assume a particular degree of correlation among the experiments is eliminated.
When performing na\"ive averages, systematic uncertainties are generally taken to be either fully correlated or completely uncorrelated among the experiments whose results are being combined.
But in reality there can be differences in how each experiment is affected.
For example, when performing a fit to an invariant mass distribution, different patterns of migration between event species in each experiment may lead to quite different effects on the signal yield when varying the
rate
of a particular background.

All of the benefits mentioned here are demonstrated by two examples from the domain of high energy physics, which are presented in Appendix~\ref{App:Bexample}.
The joint fits in the examples were realised by applying the \jfit implementation described in Sec.~\ref{sec:jfitFramework}.
In addition, it is noted that there are other positive side effects
of joint fitting, besides those that are exemplified in the appendices, arising
from
obliging collaborations to coordinate their models and conventions prior to
performing their analyses.

\section{The master-worker architecture and the \textit{J}\hspace{-0.15em}{\normalfont \normalsize{FIT}} framework}
\label{sec:jfitFramework}

In the master-worker approach, the master drives the fit by sending sets of parameters to the experiment-specific workers.
Each worker returns the value of the likelihood function to the master, which in turn updates the parameters using an optimisation routine, such as the MIGRAD algorithm of the MINUIT optimisation package~\cite{ref:minuit}, which is widely used in the field of high energy physics.
The procedure is repeated until the fit has converged.
This architecture keeps the calculation of the individual likelihood functions ${\cal L}_A$ and ${\cal L}_B$ separate, allowing a flexible treatment of any experiment-specific parameters, which can either be controlled by the master, or declared only within the corresponding worker.
In the second case, the workers perform an additional minimisation with respect to their specific parameters at each step of the procedure, keeping the common parameters fixed.
The values of the minimised functions are then returned to the master.
Performing joint fits using such an architecture has several advantages:
\begin{itemize}
\item any fitting algorithm can be readily incorporated as a worker in this scheme;
\item there is no requirement that the workers be homogeneous;
\item the data do not need to be rewritten in any external format and can be readily used;
\item experimental collaborations can keep private their data and procedures.
\end{itemize}

The \jfit framework is an open-source implementation of a master-worker architecture in the \cpp programming language.
It has been developed within the \laura package~\cite{ref:laura}, using elements from the \root data analysis framework~\cite{ref:root}.
While the main purpose of the \laura package is performing Dalitz-plot analyses in high-energy physics, \jfit can be easily used in other contexts.
The structure of the \jfit implementation is based on the following steps:
\begin{itemize}

  \item Initialisation:
	both master and workers initialise their internal structure to either
	drive the fit (master) or calculate the likelihood given a set of
	parameters (workers). 
 
  \item Synchronisation:
	the master-worker communications are handled using several classes in the \root framework.\footnote{
	This implementation is inspired by the \root macros:\\
	http://root.cern.ch/root/html/tutorials/net/authserv.C.html\\
	http://root.cern.ch/root/html/tutorials/net/authclient.C.html}
	The master starts a server, by instantiating a \texttt{TServerSocket} object with the number of the port to which it should be bound, and waits for the worker nodes to connect.
	To establish a connection, each worker instantiates a \texttt{TSocket} object by specifying the hostname and port number of the master.
	The master receives the connection in the form of another \texttt{TSocket} object, which it stores in a \texttt{TMonitor} instance.
	The connections can be secured, e.g. via ssh, if required.
	After successfully connecting, each worker awaits instructions from the master.
	There then follows an exchange of information regarding the parameters known to each worker and their initial values.
	All such communications in this and subsequent steps are conducted using instances of the \texttt{TMessage} class, which is a I/O buffer into which basic types and more complex objects can be serialised for transmission via the network sockets.

  \item Minimisation:
	the master starts the fitting procedure by sending a set of parameter
	values to the workers and waits 	for their replies.
	Upon receipt of the parameters, each worker calculates the value of its
	corresponding likelihood function and sends the result back to the master.
	The worker results are then combined by the master and given to the
	optimiser, which updates the fit parameters and the master then sends a
	new request to the workers.
	This step is repeated until convergence is achieved.
	The uncertainties on the parameters are evaluated by a dedicated procedure that uses the same logic and tools for the communication between the master and workers.

  \item Termination:
	the master returns the results of the fit and terminates the
	connections with the workers.

\end{itemize}
The master routines are implemented within the \texttt{LauSimFitMaster} class~\cite{ref:master} in the \laura package, while the communication methods of the workers are implemented in the \texttt{LauSimFitSlave} abstract base class~\cite{ref:slave}.
The main tasks of the master are to set up and provide the appropriate information to the minimiser, to keep track of which fit parameters are associated with which of the workers, and to communicate with the workers so as to delegate the calculation of the likelihood and other pre- and post-processing tasks.
The worker communication methods respond to the various messages from the master and call the appropriate pure virtual member function, the implementation of which in the concrete derived classes should then perform the required actions.

As a consequence, a large fraction of code developed by each experiment to perform dedicated fits can be readily reused in this scheme; it requires only the addition of a class that inherits from \texttt{LauSimFitSlave}.
This class should implement the functions that are pure virtual in the base class in such a way as to call the appropriate parts of the pre-existing code to perform the actions required at each stage of the fit (e.g., calculating the likelihood value).
Internal changes in one experiment (e.g. new data format or additional specific parameters) require only a modification of this class and are otherwise completely transparent to the framework.
More specifically, to write a class that inherits from \texttt{LauSimFitSlave} involves writing the implementation of the following eight pure virtual member functions:
\begin{enumerate}
	\item \texttt{initialise}: perform any actions required to ensure that the fit model is ready to be applied to the data;
	\item \texttt{verifyFitData}: read the data to be fitted and verify that all required variables are present; the name of the data file and (optionally) the name of a particular structure within the file, e.g. a \root tree, will have been provided by the user when calling the \texttt{LauSimFitSlave::runSlave} function and they are passed on as arguments to this function;
	\item \texttt{prepareInitialParArray}: the fit parameters that are to be varied in the fit should be inserted, in the form of \texttt{LauParameter} objects, into the \texttt{TObjArray} that is the argument to the function;
	\item \texttt{readExperimentData}: read the data for the current experiment into memory;
	\item \texttt{cacheInputFitVars}: allow the fit model to cache any information that it can from the data;
	\item \texttt{setParsFromMinuit}: update all fit parameters with the new values provided by the minimiser;
	\item \texttt{getTotNegLogLikelihood}: calculate the negative log likelihood;
	\item \texttt{finaliseExperiment}: store the fit results, in particular the fit status, negative log likelihood, covariance matrix and final parameter values and uncertainties; perform any necessary post-processing (e.g. rotation of phases into a particular interval) of the parameters and then return them to the master.
\end{enumerate}
An example of such a derived class, which uses the \roofit framework to build the fit model and evaluate the likelihood, is shown in Appendix~\ref{sec:roofit-slave}.

It is important to note that the \jfit framework can also be used to perform simultaneous fits to different datasets within an experiment.
For example, it has been used in Ref.~\cite{ref:Bu2DKpi} to account for the different variation over phase space of the signal reconstruction efficiency in different trigger categories.
In addition, it has been used in Ref.~\cite{ref:Bd2DKpi-gamma} to enable the extraction of both \CP-violation and hadronic parameters from a simultaneous fit to multiple decay channels.

The overhead introduced by the \jfit framework itself is small.
For instance, in the example in Appendix~\ref{DPexample} the mean time for each individual fit to be performed was 33 seconds (with an RMS of 6 seconds), while the joint fits using \jfit took 35 seconds (with an RMS of 5 seconds).

\section{Conclusions}
\label{sec:conclusion}

In this paper, we have presented \jfit: a software framework for obtaining combined experimental results through joint fits of datasets from several experiments.
The primary goal of the \jfit framework is to permit experimental collaborations to straightforwardly perform joint fits by allowing them to plug in their existing fitting software and to use their data in its original format.
It is implemented in the \laura Dalitz-plot analysis package~\cite{ref:laura}, using the network communication classes (\texttt{TMessage}, \texttt{TMonitor}, \texttt{TServerSocket} and \texttt{TSocket}) from the \root framework.
The use of \jfit is not limited to the context of Dalitz-plot analyses.

Advantages of joint fitting have been discussed.
Correctly accounting for correlations between parameters in likelihood functions in different experiments, which is an intrinsic property of such fits, results in improved combinations, and more reliable uncertainties.
Thus, these fits provide a means to better exploit data from multiple experiments.
The \jfit framework, based on a master-worker architecture, allows joint fits to be performed keeping the data separated and using independent, heterogeneous fitting programs.
It simplifies the process with respect to data access policies and allows existing code to be reused with minimal changes, thus saving resources.

\section*{Acknowledgments}

The work presented in this paper was started in the purpose of performing joint analyses between the \babar\ and Belle collaborations.
We would like to thank Fran\c{c}ois Le Diberder both for initiating and generating interest in this project among the \babar\ and Belle collaborations and the authors of \root, and for fruitful discussions concerning this paper.
We would also like to thank Roger Barlow for his extremely valuable advice and comments regarding the statistical aspects and examples, and Tim Gershon and Matthew Charles for their very helpful input on the manuscript.
This work is supported by the Science and Technology Facilities Council (United Kingdom), the European Research Council under FP7, and the US Department of Energy under grants DE-FG02-92ER40701 and DE-SC0011925.

\clearpage

\section*{Appendix}
\appendix
\section{Maximum-likelihood estimation}
\label{max}

Maximum-likelihood estimation is a widely-used method of fitting parameters of a 
model to some data and providing an estimate of their uncertainties.
Here we briefly review this technique; more details can be found in
statistics textbooks, and, for example, in Ref.~\cite{ref:pdg}.

Consider a set of $N$ independent observations $x_1,...,x_N$ of a random variable $x$ 
following a probability density function modelled by ${\cal P}(x;\theta)$, where $\theta$ 
denotes the unknown model parameter(s) that should be estimated. Both $x$ and
$\theta$ can be multidimensional. The likelihood 
function $\cal{L}$ is defined as:
\begin{equation}
{\cal L}(\theta) = \prod_{i=1}^{N} {\cal P}(x_i;\theta).
\end{equation}
Since the $N$ observations $x_i$ are given, the likelihood is only a function of $\theta$. 
The resulting estimate $\hat{\theta}$ of the parameter $\theta$ is defined as the value that maximises the likelihood, i.e.
\begin{equation}
\hat{\theta} = {\max}_{\theta}{\cal L}(\theta).
\end{equation}
It is often convenient to minimise the negative logarithm of the likelihood (NLL) instead of maximising the likelihood, which yields the same value of $\hat{\theta}$.
If the number of observations is random, an additional term is usually included in the likelihood to form the so-called extended likelihood:
\begin{equation}
{\cal L}_{\rm ext}(\theta) = \frac{e^{-n}n^N}{N!}\prod_{i=1}^{N} {\cal P}(x_i,\theta),
\end{equation}
where $n$ denotes the expected number of events.
Obtaining combined measurements from several datasets is achieved by maximising the product of likelihoods from the different datasets, as discussed in Sec.~\ref{sec:JointFitsAndExamples}.

The minimisation of the NLL can be, and usually is, done numerically. Among the 
packages available to perform this task, MINUIT~\cite{ref:minuit} is one of the most popular in the 
field of high energy physics. Its default algorithm, MIGRAD, is based on a 
variable-metric method that computes the value of the function and its gradient 
at each step of the procedure.
If the functional form of the derivative is not, or cannot be supplied by the user, the gradient is evaluated by finite differences.
The minimisation stops when the difference of the value of the function between two 
successive steps reaches a specific threshold. Parabolic uncertainties on the 
parameters are estimated by inverting the matrix of second derivatives evaluated at 
the minimum of the function. 
In the case of non-Gaussian likelihood, the uncertainties are estimated
by an algorithm that scans the likelihood for each parameter separately, minimising 
the likelihood each time with respect to the remaining parameters.

\clearpage

\section{Examples demonstrating the benefits of joint fitting}
\label{App:Bexample}

All the fits performed within the two examples described in this appendix have been realised by applying the \jfit framework.

\subsection{Resonance mass and rate measurement}
\label{ResExample}

As a first example, we examine the problem of combining the results of two experiments that found a resonance in an invariant mass spectrum, at a mass of $126\gevcc$.
The parameters of interest are:
\begin{itemize}
	\item
	the mass of the resonance, $m_{\rm Res}$;
	\item
	its observed rate in the final state under study, which results from the product of the production cross section, $\sigma$, and the decay branching fraction, \BR, normalised to its expected value, e.g., that predicted by a particular model, $R = \frac{\sigma\times\BR}{\sigma_\mathrm{Exp.}\times\BR_\mathrm{Exp.}}$.
\end{itemize}
These are obtained from a one-dimensional maximum-likelihood fit to the distribution of invariant mass, $m$, of the final-state particles.

\begin{figure}[htbp]
\centering
\includegraphics[width=0.55\textwidth]{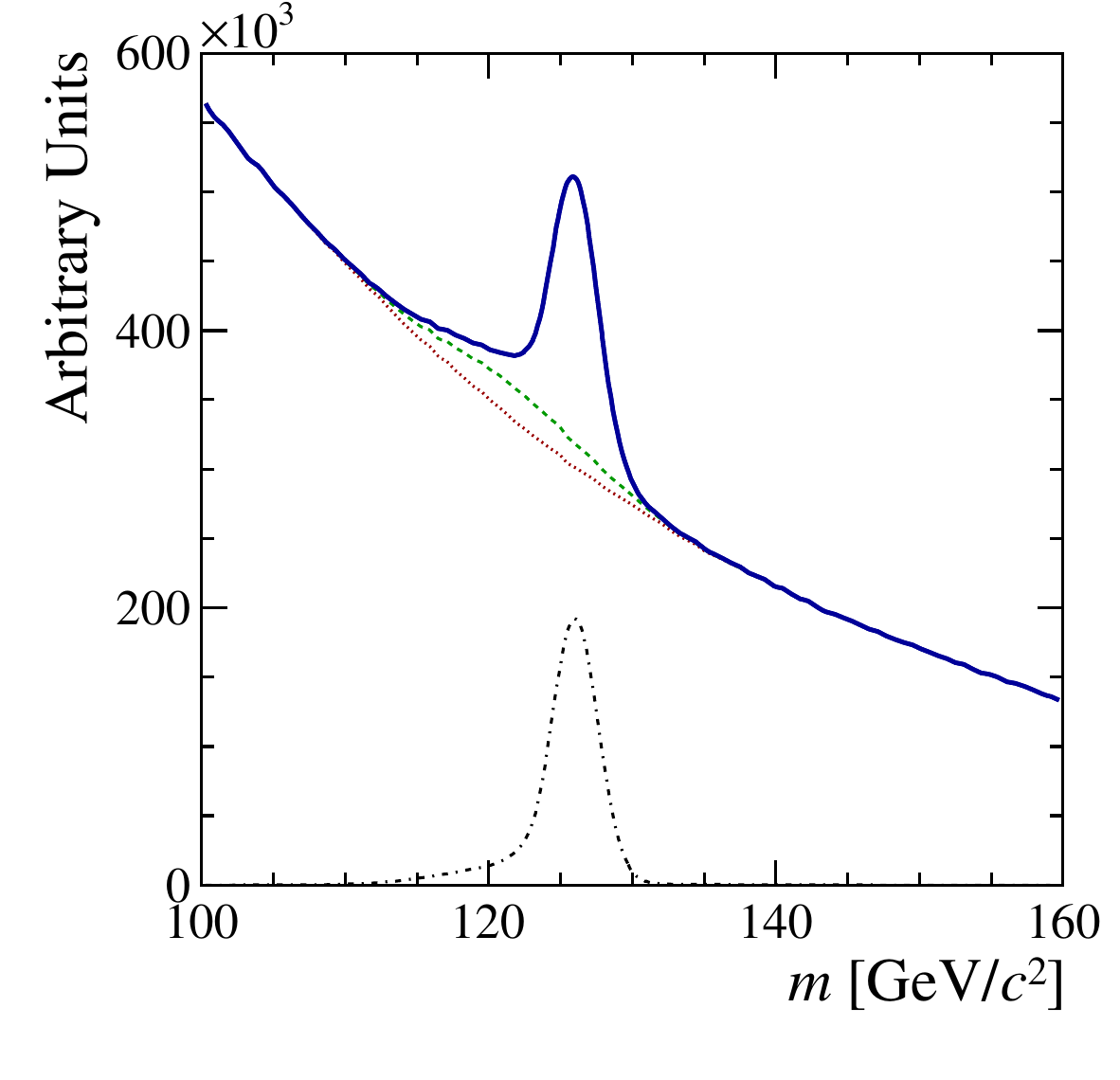}
\vspace{-0.5 em}
\caption{\small
The invariant-mass distribution of a sample generated for experiment 1. The solid (blue) curve shows the full sample. The dash-dotted (black) curve corresponds to the distribution of signal events, while the dotted (red) and the dashed (green) curves show the distributions for the combinatorial and total background, respectively. 
\label{fig:massPlot}}
\end{figure}

Samples corresponding to the datasets of experiments 1 and 2 are generated in the invariant-mass range $[100,160]\gevcc$ according to a model containing three events species: signal, peaking background and combinatorial background.
Signal events are generated from a sum of two Crystal Ball (CB) functions~\cite{ref:CB}: core and tail, with different peak positions.
Peaking and combinatorial background events are generated from a Gaussian and a fourth order Chebyshev polynomial function, respectively.
This model is roughly inspired by the search for the Standard Model Higgs boson from ATLAS~\cite{ref:atlas} and CMS~\cite{ref:cms}, although, with rather different event yields and signal to background ratios.
For simplicity, the same signal and background models are used to generate events for both experiments.
Hence, both the underlying physics and some experimental factors, such as the resolution, are assumed to be unique.
However, to emulate in a simple way different overall efficiencies and event-selection strategies, the datasets corresponding to the two experiments have significantly different numbers of signal and background events.
In both cases, the mean value of the signal yield corresponds to $R=1.0$.
Five hundred pairs of datasets are generated, where the numbers of events of each species are Poisson-distributed, using probability density functions (PDFs) that are summarised in Table~\ref{tab:genModels}.
The invariant mass distribution of experiment 1 is shown in Fig.~\ref{fig:massPlot}.

\begin{table}[htbp]
\caption{\small
Summary of the model used to generate events.
The average number of events of each species generated per experiment is denoted $N$.
The peak positions and Gaussian widths of the CB and Gaussian functions are denoted $\mu$ and $\sigma$, respectively, with the superscript core or tail, and with the subscript CB, as appropriate.
The CB tail parameters are $\alpha$ and $n$.
The coefficient of the $i^{\rm th}$ power term in the polynomial is denoted $c_i$.
\label{tab:genModels}}
\renewcommand{\arraystretch}{1.1}
\centering
\begin{tabular}{ |l|l|l|c| }
\hline
%%%%%%%%%%%%%%% Generagion
%
Event species & Function & Parameter & Value\\ \hline
\multirow{11}{2.7cm}{Signal} & \multirow{11}{2.7cm}{Sum of two CB functions (core and tail)} & $N$ (experiment 1) & 3000\\
& & $N$ (experiment 2) & 10000\\
& & $\mu_{\rm CB}^{\rm core}$ & 126\gevcc \\
& & $\sigma_{\rm CB}^{\rm core}$ & 1.6\gevcc \\
& & $\alpha^{\rm core}$ & 1.5 \\
& & $n^{\rm core}$ & 3.0 \\
& & Fraction of tail & 10\% \\
& & $\mu_{\rm CB}^{\rm tail}$ & 120\gevcc \\
& & $\sigma_{\rm CB}^{\rm tail}$ & 4.0\gevcc \\
& & $\alpha^{\rm tail}$ & $-1.0$ \\
& & $n^{\rm tail}$ & $\phantom{-}4.0$ \\ \hline
\multirow{4}{2.7cm}{Peaking background} & \multirow{4}{2.8cm}{Gaussian} & $N$ (experiment 1) & 1000\\
& & $N$ (experiment 2) & 7500\\
& & $\mu$ & 122\gevcc \\
& & $\sigma$ & 5.0\gevcc \\ \hline
\multirow{6}{2.7cm}{Combinatorial background} & \multirow{6}{2.8cm}{\nth{4} order Chebyshev polynomial} & $N$ (experiment 1) & 60000\\
& & $N$ (experiment 2) & 500000\\
& & $c_1$ & $-0.682$ \\
& & $c_2$ & $\phantom{-}0.122\,[\!\gevcc]^{-1}$ \\
& & $c_3$ & $-0.013\,[\!\gevcc]^{-2}$ \\
& & $c_4$ & $-0.003\,[\!\gevcc]^{-3}$ \\ \hline
\end{tabular}
\renewcommand{\arraystretch}{1.0}
\end{table}

The maximum-likelihood fits are performed in the same invariant-mass range, $[100,160]\gevcc$.
Again, the same model is used for the two experiments.
However, the fitted model differs slightly from that used to generate events in order to account for the lack of knowledge of the underlying physics.
The invariant mass of the signal is modelled by the sum of a Crystal Ball function, describing the core distribution, and a Gaussian describing the tails.
Backgrounds are modelled by the same functional forms used to generate events.
The signal peak position and all the combinatorial background parameters are varied in the fit, as well as the yields of the three events species.
All the other parameters are fixed.
The PDFs used in the fits are summarised in Table~\ref{tab:fitModels}.

\begin{table}[htbp]
\caption{\small
Summary of the model used to fit events.
Parameter notations are the same as in Table~\ref{tab:genModels}.
Values of fixed parameters are given in the ``Value'' column; ``Gen.'' means that the fixed value is identical to that used for generation.
The parameters $\mu^{\rm tail}$ and $\mu_{\rm CB}^{\rm core}$ are constrained to take the same value.
\label{tab:fitModels}}
\renewcommand{\arraystretch}{1.1}
\centering
\begin{tabular}{ |l|l|l|c| }
\hline
%%%%%%%%%%%%%%% Fit
%
Event species & Function & Parameter & Value\\ \hline
\multirow{7}{2.7cm}{Signal} & \multirow{7}{2.8cm}{Sum of a core CB function and a Gaussian tail} & $N$ & varied\\
& & $\mu_{\rm CB}^{\rm core}$ & varied\\
& & $\sigma_{\rm CB}^{\rm core}$ & 1.5\gevcc \\
& & $\alpha^{\rm core}$, $n^{\rm core}$ & Gen. \\
& & Fraction of tail & Gen. \\
& & $\mu^{\rm tail}$ & $=\mu_{\rm CB}^{\rm core}$\\
& & $\sigma^{\rm tail}$ & 3.8\gevcc \\ \hline
\multirow{3}{2.7cm}{Peaking background} & \multirow{3}{2.8cm}{Gaussian} & $N$ & varied\\
& & $\mu$ & Gen. \\
& & $\sigma$ & Gen. \\ \hline
\multirow{3}{2.7cm}{Combinatorial background} & \multirow{3}{2.8cm}{\nth{4} order Chebyshev polynomial} & $N$ & varied\\
& & &  \\
& & $c_1$, $c_2$, $c_3$, $c_4$ & varied \\ \hline
\end{tabular}
\renewcommand{\arraystretch}{1.0}
\end{table}

\begin{figure}[htbp]
\centering
\includegraphics[width=0.44\textwidth]{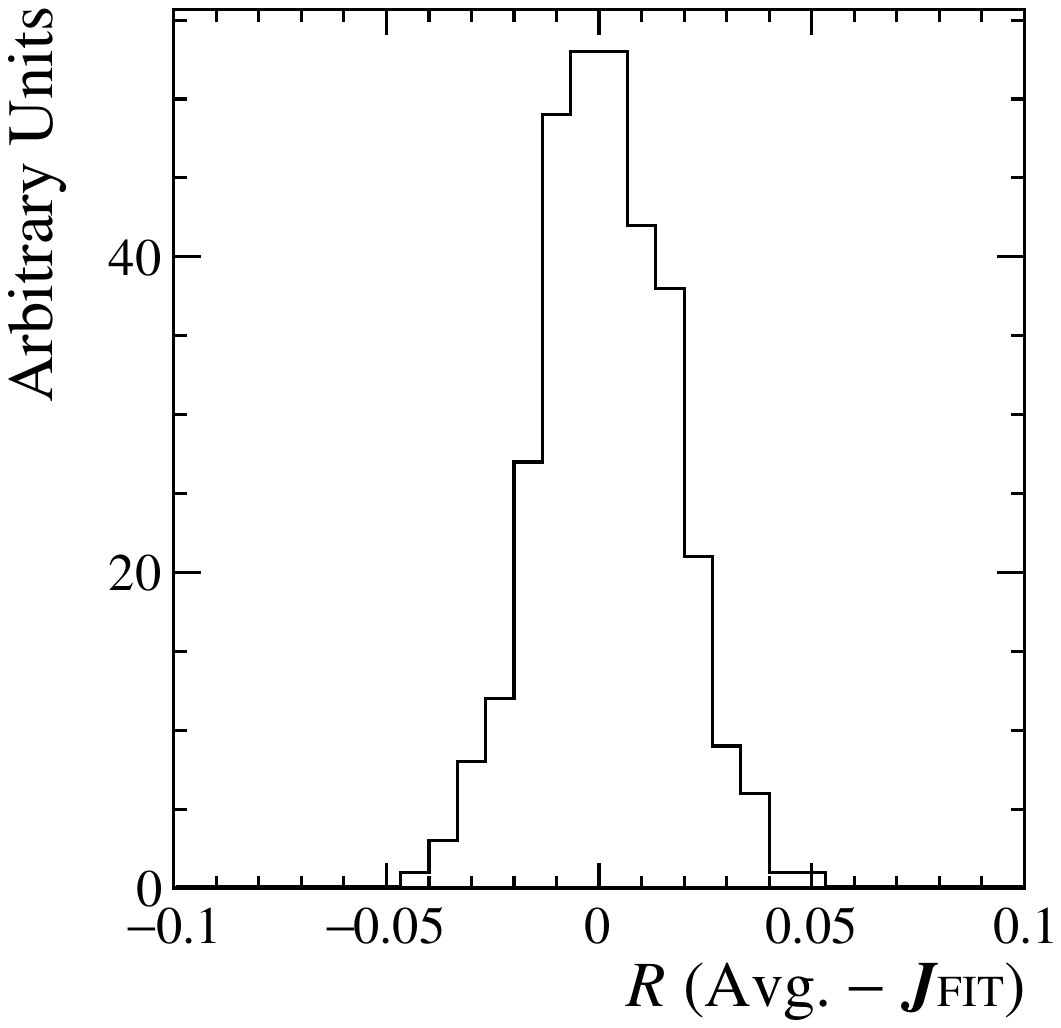}
\includegraphics[width=0.44\textwidth]{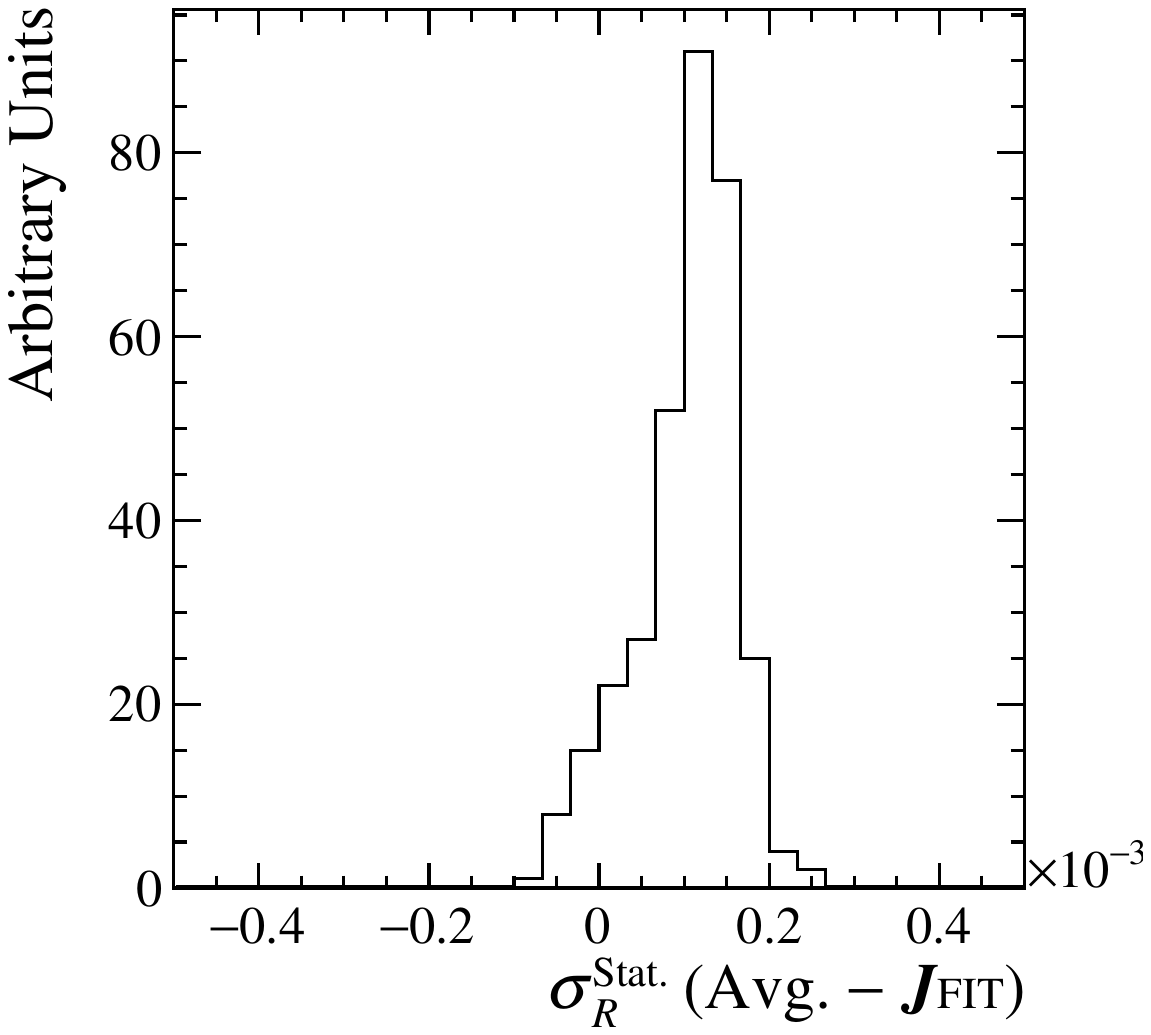}
\caption{\small
The distribution of the difference between the results of combinations performed by na\"ive averaging and by joint fits performed in the \jfit framework, for the (left) central values of the resonance rate, and the (right) corresponding statistical uncertainties.
The former shows that in this particular case there is no difference in the bias induced by the two methods, and the latter shows that the statistical uncertainty (typically 0.027) is larger by $\sim 1\%$ in na\"ive averages.
\label{fig:ResBFNaiveVsJoint}}
\end{figure}

After fitting each of the individual samples, we obtain combined results for the mass of the resonance and its rate by two different methods:
\begin{enumerate}
\item a na\"ive average of the individual results, taking into account the parabolic uncertainties of the fits to the individual samples;
\item a joint fit to the two samples in the \jfit framework,
using the full likelihood function.
\end{enumerate}
As the likelihood functions are fairly Gaussian, asymmetric uncertainties have not been considered for the na\"ive averaging.
The typical statistical uncertainty obtained by the two methods is $0.027$ (corresponding to a $\sim 3\%$ relative uncertainty) for the normalised rate and $0.054\gevcc$ for the mass of the resonance.
Moreover, no bias has been observed in the extraction of the mass of the resonance by the two methods, while they both show a negative bias in the extraction of the rate, as expected from the different signal models used to generate and fit the data samples: the former has a wider signal peak than the latter. The results of the two methods for the resonance rate are compared in Fig.~\ref{fig:ResBFNaiveVsJoint}.
In this particular case, the bias induced by the two methods is similar and thus the central values are in good agreement.
There is a small effect on the statistical uncertainty, which is very slightly larger in na\"ive averages.
This behaviour may differ, in size and direction, in other cases.
However, as the joint fits correctly take into account all correlations, they provide more reliable results.

\begin{figure}[htbp]
\centering
\includegraphics[width=0.41\textwidth]{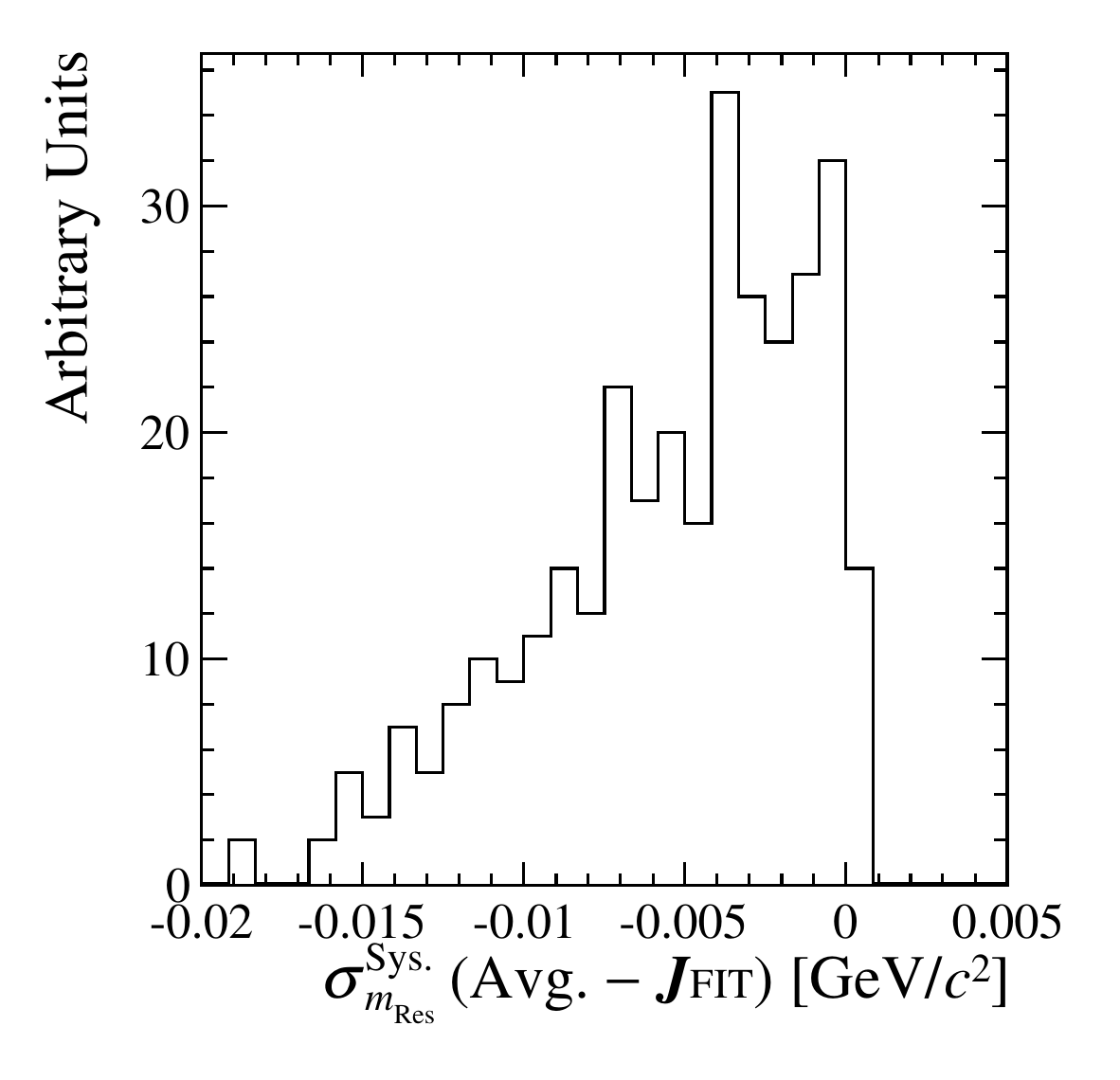}
\includegraphics[width=0.41\textwidth]{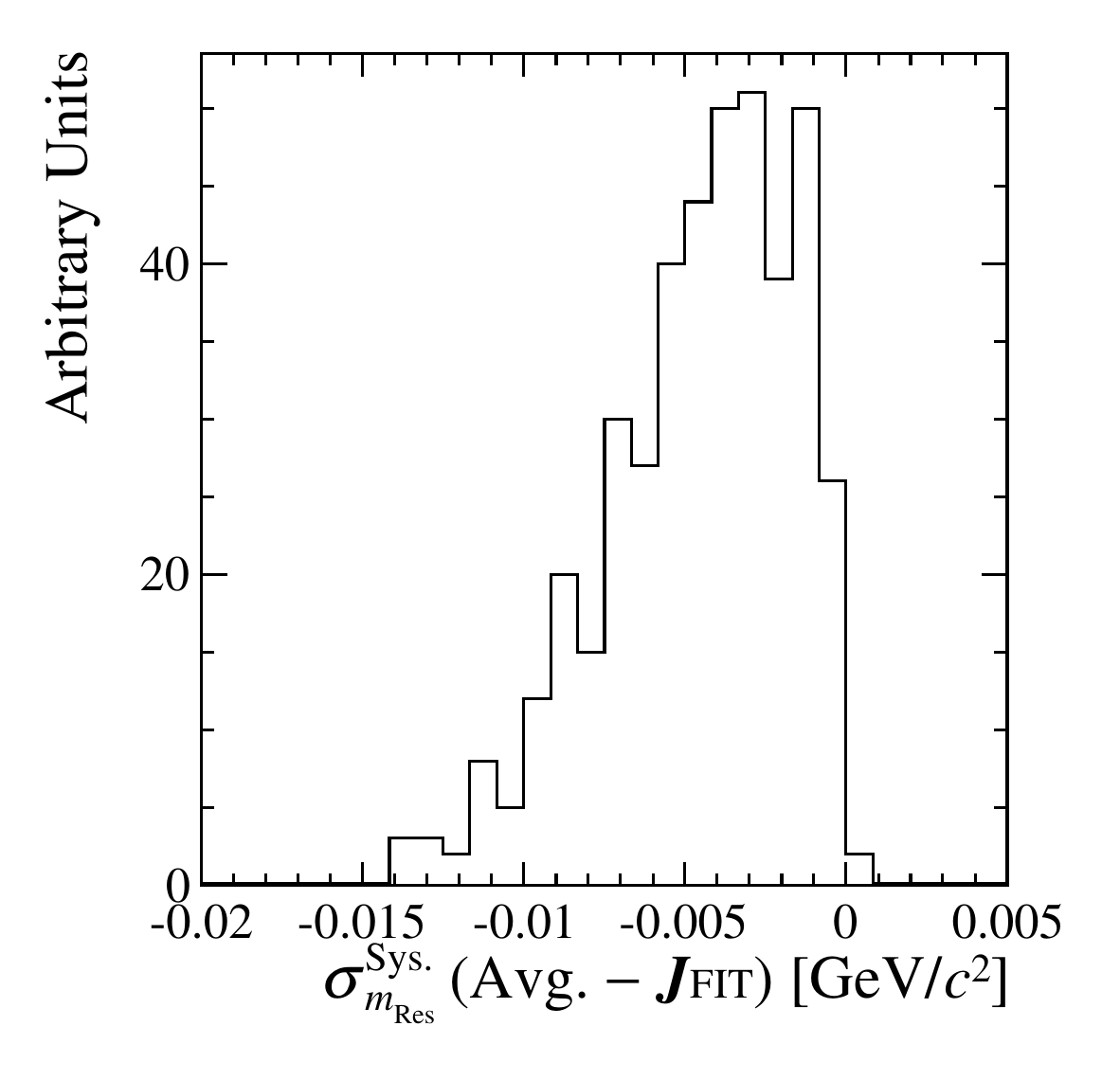}
\includegraphics[width=0.41\textwidth]{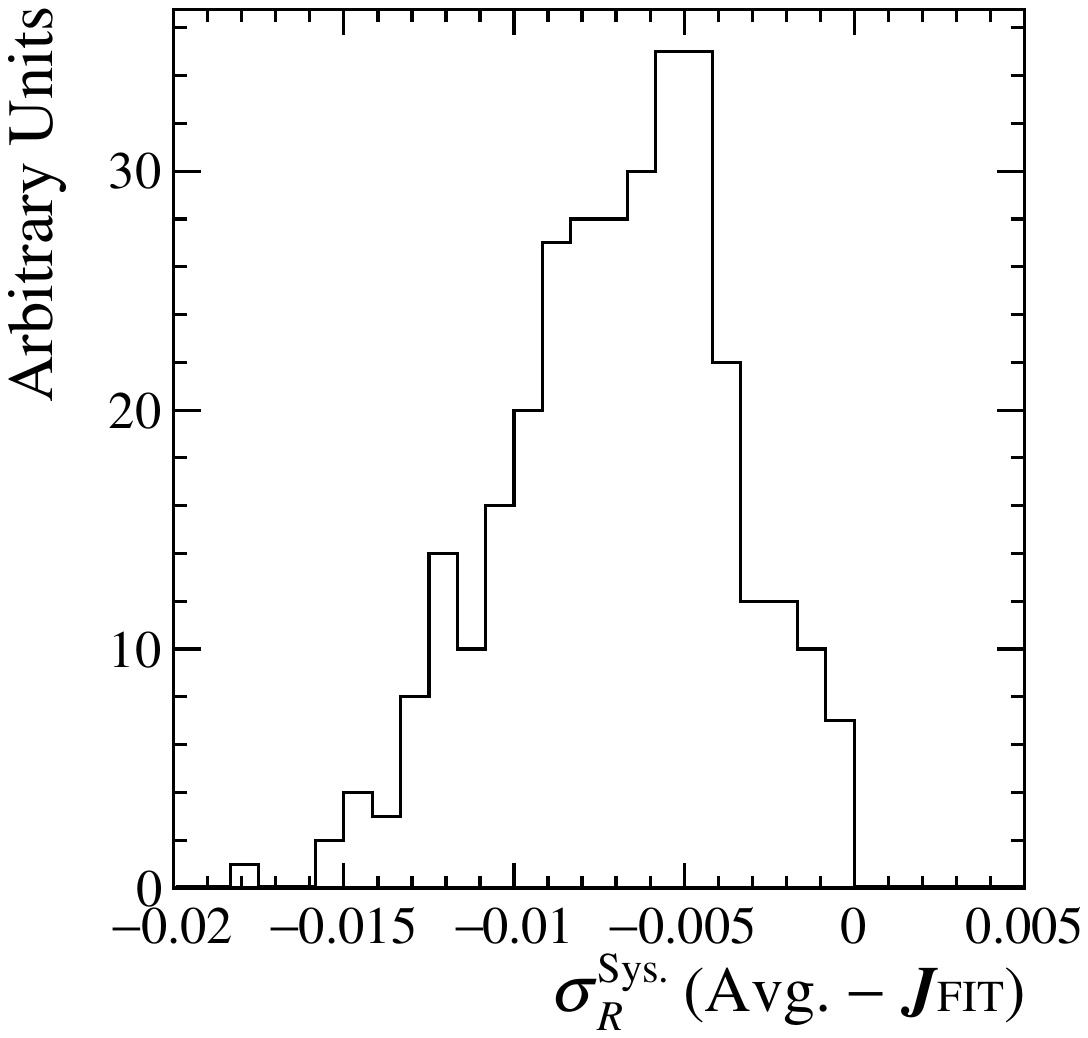}
\includegraphics[width=0.41\textwidth]{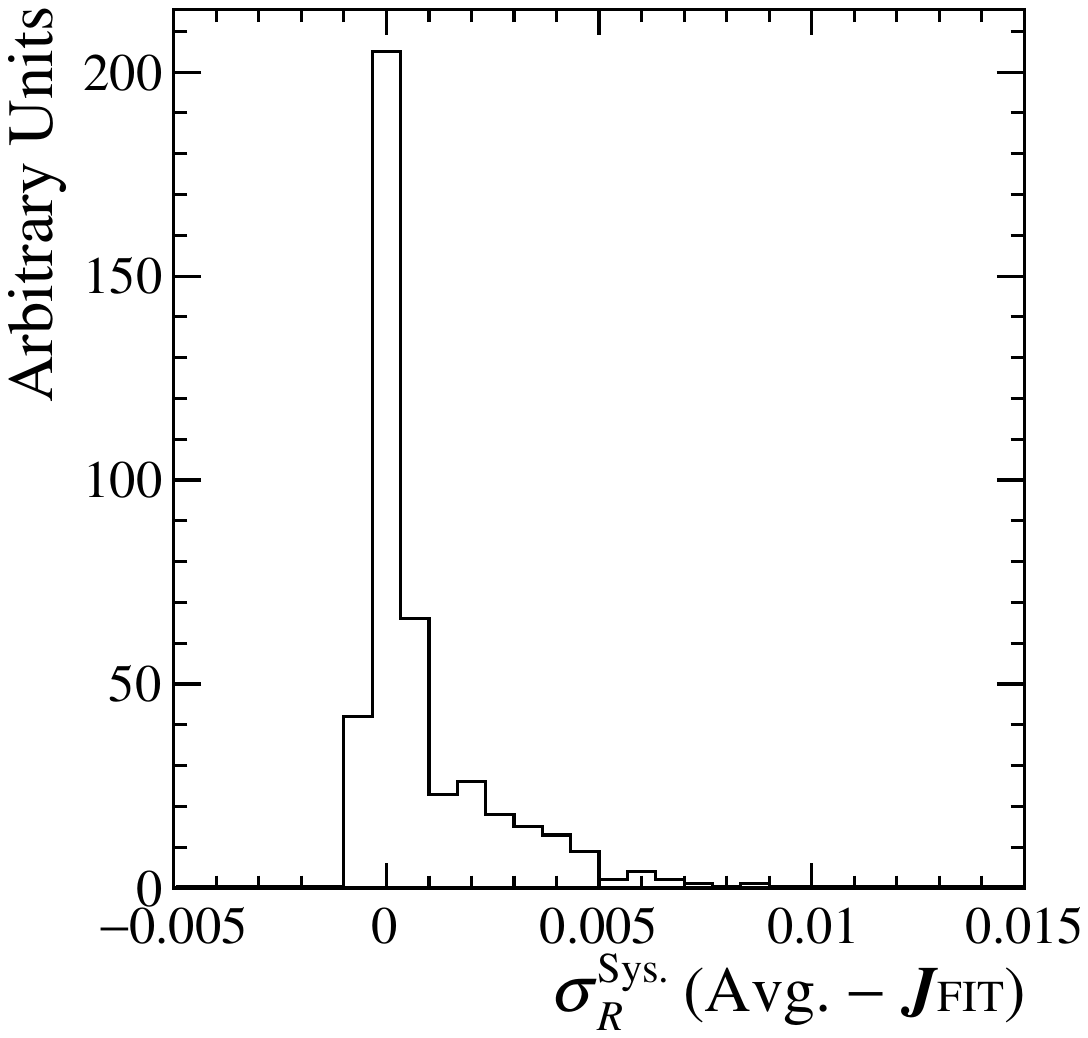}
\caption{\small
The difference between the systematic uncertainty values obtained in na\"ive averages and in joint fits performed in the \jfit framework.
The systematic uncertainty considered is that arising from the width of the peaking background component.
The na\"ive averages assume that the systematic effects are 100\% correlated between the two experiments.
The effect on both (top) $m_{\rm Res}$ and (bottom) the normalised rate, is shown for a peaking background Gaussian width of (left) 3.0\gevcc and (right) 7.0\gevcc.
In most of the cases the uncertainty is underestimated by na\"ive averaging.
The exception is the uncertainty on the rate obtained by increasing the width to 7.0\gevcc (bottom right plot).
The typical statistical uncertainty on the resonance mass is 0.054\gevcc, while that of the rate is 0.027.
\label{fig:massAndBFsyst}}
\end{figure}

We expect larger differences between the two combination methods to arise when systematic uncertainties are evaluated.
Parameters of the peaking background are considered to be badly known and, as such, to be sources of systematic uncertainties.
To evaluate the effect from the width of the distribution, the corresponding PDF parameter is fixed to values $2\gevcc$ above and below its nominal value, namely 3.0\gevcc and 7.0\gevcc.
The variations of the resulting resonance mass and rate are considered as systematic uncertainties.
The same procedure is applied in individual and joint fits.
Figure~\ref{fig:massAndBFsyst} shows a comparison between the systematic uncertainties obtained in na\"ive averages, where the individual systematic effects are considered to be 100\% correlated, and in joint fits.
Differences are due to the fact that correlations are correctly taken into account in joint fitting.
The comparison shows clearly that the effect on systematic uncertainties can be large (in some cases around 50\% of the statistical uncertainty or 10\% of the systematic uncertainty), and they may be either underestimated or overestimated by na\"ively averaging the results.

\clearpage

\subsection{Amplitude analysis of a hadronic three-body \B-meson decay}
\label{DPexample}
A second example is from the domain of flavour physics. It involves a larger number of parameters in the fit, and illustrates other advantages of joint fitting.
For simplicity, we consider signal events only.

A key part of the physics programme of the \B-factories, \babar\ and Belle, consisted of amplitude (Dalitz-plot) analyses of 3-body \B-meson decays~\cite{ref:pbf}.
More recently such studies are performed by the LHCb experiment (see, for example, Refs.\cite{ref:lhcb_JpsiKK, ref:lhcb_DKpi, ref:lhcb_KSpipi}), and, in the near future, will be also undertaken by Belle-II.
These analyses provide measurements of CKM angles and access to observables sensitive to physics beyond the Standard Model of particle physics as well as information on the resonant structure of decays.

In general, Dalitz-plot analyses are limited by the sample size, and they usually have a strong dependence on model assumptions. Due to these characteristics, a joint analysis, profiting in a coherent way of all the available data, could be a particularly fruitful approach compared to a simple combination of results from separate analyses.
One of the first steps in Dalitz-plot analyses consists of determining which
resonant or non-resonant intermediate states should be included as components of the signal model. It is important to notice that minor, poorly determined signal components are a major source of the so called “model uncertainty” that is often a large systematic effect in such analyses. A joint analysis provides a more powerful determination of the components to be included in the signal model, and allows setting better limits on minor components. Another source of model dependence comes from the choice of particular parameterisations of intermediate decay modes in the signal model (e.g., resonance lineshapes and phase conventions). 
The fact that different collaborations often use different parameterisations can lead to difficulties in comparing their results.
In some cases a direct comparison of such results can be less meaningful; they are less useful for the community, and averaging them becomes non trivial. Besides the benefit of grouping the expertise of the different collaborations, the coordination of signal models, which is a {\it sine qua non} for a joint fit, is therefore beneficial.
These advantages of joint fits are not explicitly illustrated in this work.

To exemplify direct advantages of joint fitting, we use the result of the \babar\ Dalitz-plot analysis of \BtoKPP\ decays~\cite{ref:kpipi}.
This analysis provided \CP-averaged branching fractions and direct \CP asymmetries for intermediate resonant and non-resonant contributions.
It reported evidence for direct \CP violation in the decay $B^\pm \to \rhoI K^\pm$, with a \CP-violation parameter $A_{\CP}=(44 \pm 10 \pm 4^{+5}_{-13})\%$, where the first quoted uncertainty is statistical, the second is systematic, and the third is the model uncertainty mentioned above.
The Belle collaboration also reported evidence of direct \CP\ violation in the same decay mode~\cite{ref:BelleKpipi} with a similar significance.
In such a situation, being able to obtain a combined result is strongly motivated.

In the \babar\ analysis, the contributions of the different intermediate states in the decay were obtained from a maximum-likelihood fit of the distribution of events in the Dalitz plot formed from the two variables $\mACSq \equiv m_{\Kpm\pimp}^2$ and $\mBCSq \equiv m_{\pipm\pimp}^2$.
As in many other Dalitz-plot analyses, the total signal amplitudes $A$ and $\overline{A}$ for \Bp\ and \Bm\ decays, respectively, were given in the isobar formalism, by
\begin{eqnarray}
A = A(\mACSq,\mBCSq) &=& \sum_j c_j F_j(\mACSq,\mBCSq) \\
\overline{A} = \overline{A}(\mACSq,\mBCSq) &=& \sum_j \overline{c}_j \overline{F}_j(\mACSq,\mBCSq) \, ,
\end{eqnarray}
where $j$ is a given intermediate decay mode.
The distributions $F_j \equiv \overline{F}_j$ are the lineshapes (e.g., Breit--Wigner functions) describing the dynamics of the decay amplitudes, and the complex coefficients $c_j$ and $\overline{c}_j$ contain all the weak-phase dependence and are measured relative to one of the contributing channels.
They were parameterised as
\begin{eqnarray}
c_j &=& (x_j + \Delta x_j) + i (y_j + \Delta y_j) \\ \nonumber
\overline{c}_j &=& (x_j - \Delta x_j) + i (y_j - \Delta y_j) \, ,
\end{eqnarray}
where $\Delta x_j$ and $\Delta y_j$ are \CP-violating parameters.

We generate $100$ signal-only datasets from the results of this analysis.
The sample size is Poisson-distributed with an expected value of $4585$, which is the signal yield obtained in the fit to the \babar\ data~\cite{ref:kpipi}.
We then consider each of the $4950$ possible pairwise combinations of these samples as datasets from two different experiments.
For the purpose of this example we focus on one of the parameters of interest of the \babar\ analysis: the \CP violating parameter $\Delta x$ of the $\rhoI K^{\pm}$ contribution, $\dxrho$.
The value used to generate events is that measured by \babar, namely $-0.160\pm0.049\pm0.024^{+0.094}_{-0.013}$.
After fitting each of the individual samples with the model used for event generation, we obtain combined results for $\dxrho$ by three different methods:
\begin{enumerate}
\item a na\"ive average of the individual results for $\dxrho$, taking into account the parabolic uncertainties of the fits to the individual samples;
\item a na\"ive average, taking into account the asymmetric uncertainties of the individual fits\footnote{The asymmetric uncertainties are obtained from the MINOS routine of the MINUIT package.}, following the prescription from Ref.~\cite{ref:pdg} (denoted $\dxrhoavg$);
\item a joint fit to the two samples in the \jfit framework,
using the full likelihood function (denoted $\dxrhojfit$).
\end{enumerate}

\begin{figure}[htbp]
\includegraphics[width=0.5\textwidth]{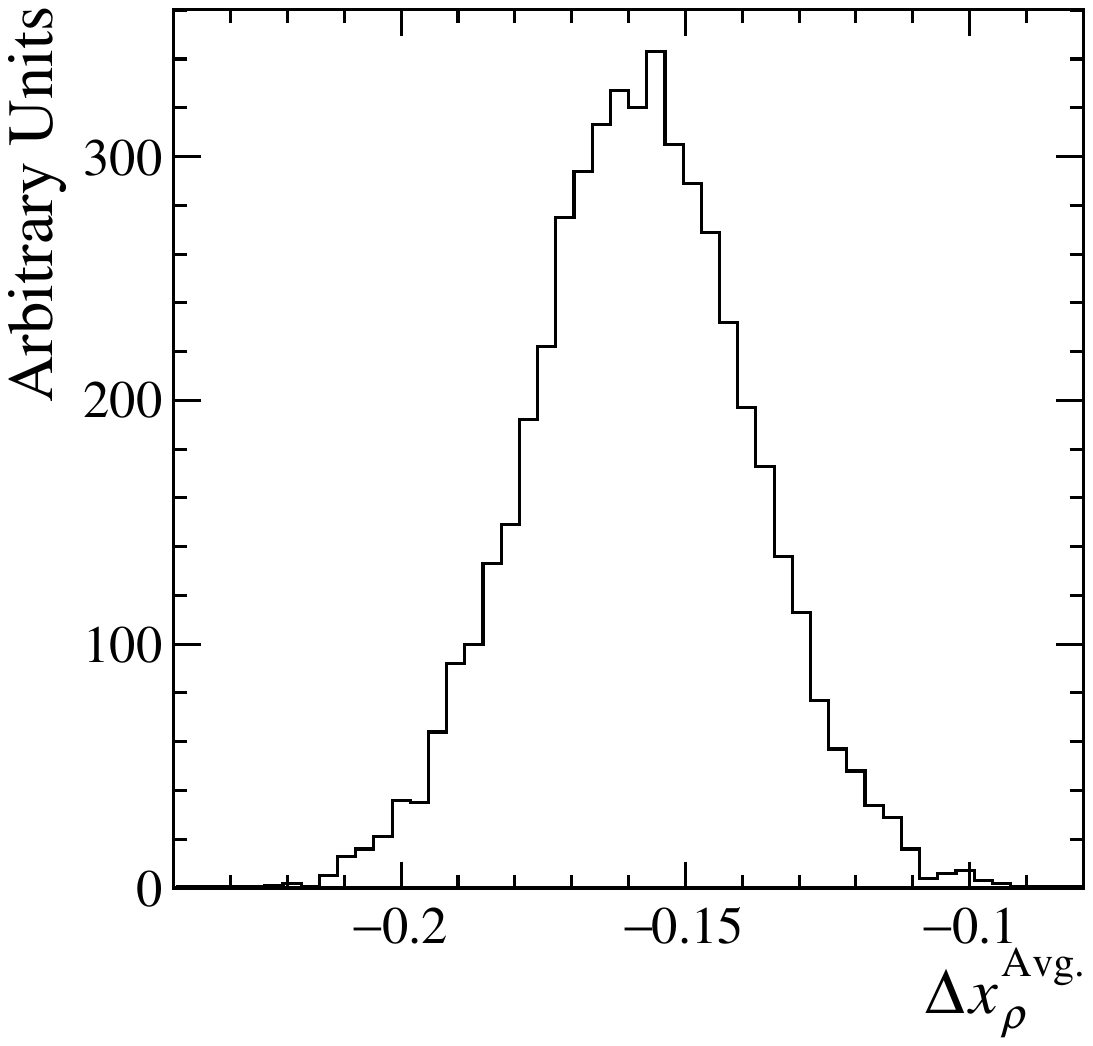}
\includegraphics[width=0.5\textwidth]{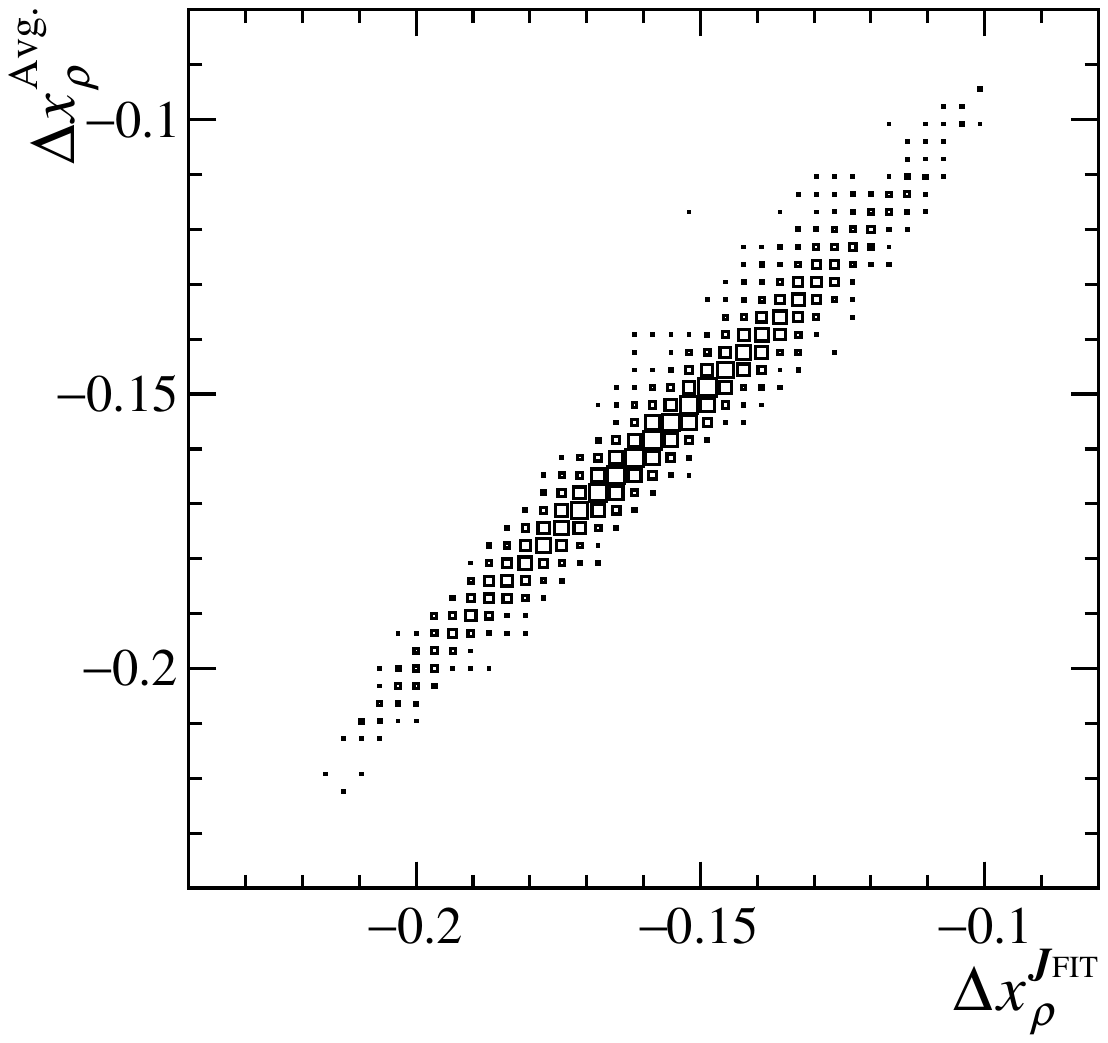}
\includegraphics[width=0.5\textwidth]{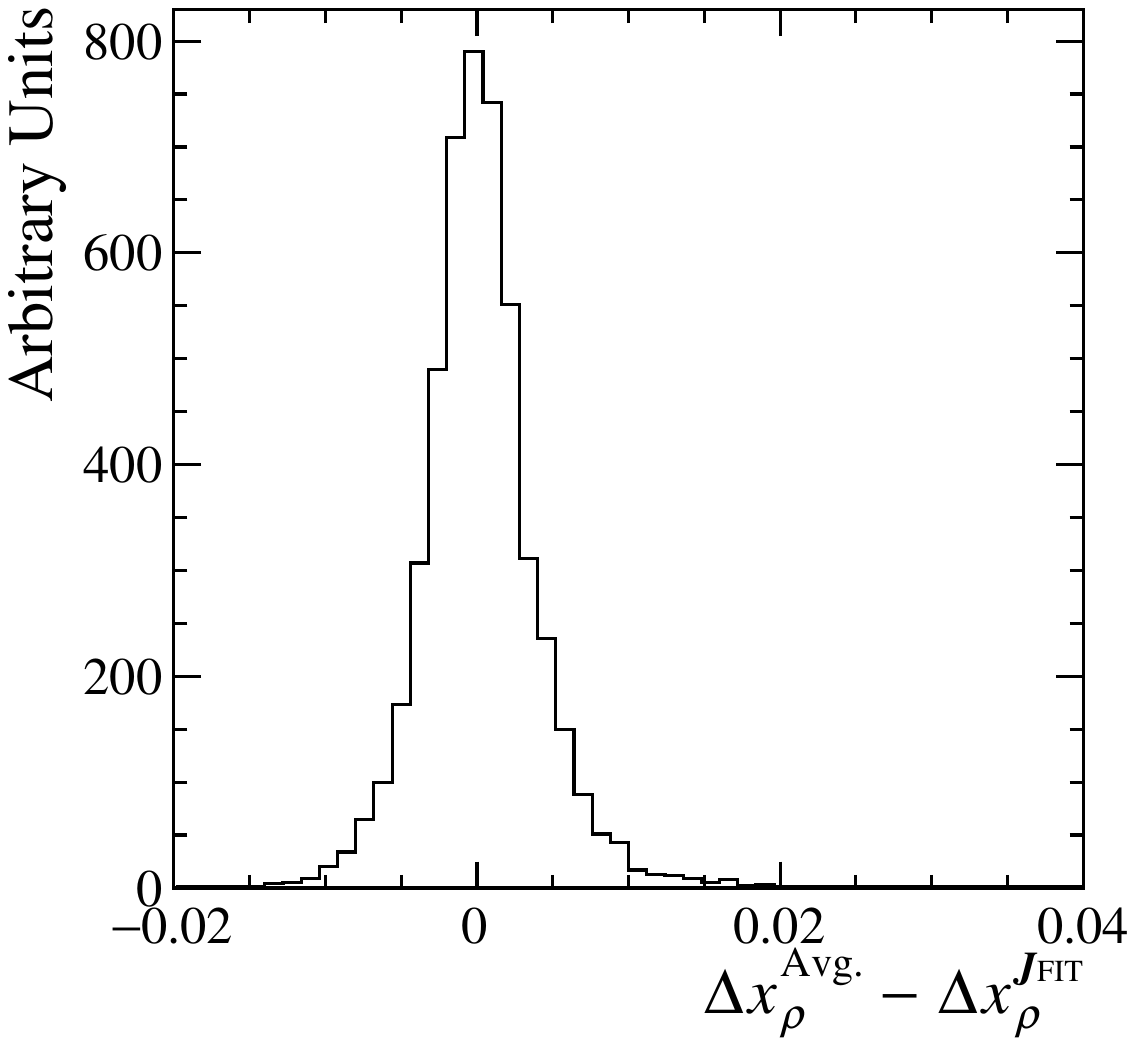}
\includegraphics[width=0.5\textwidth]{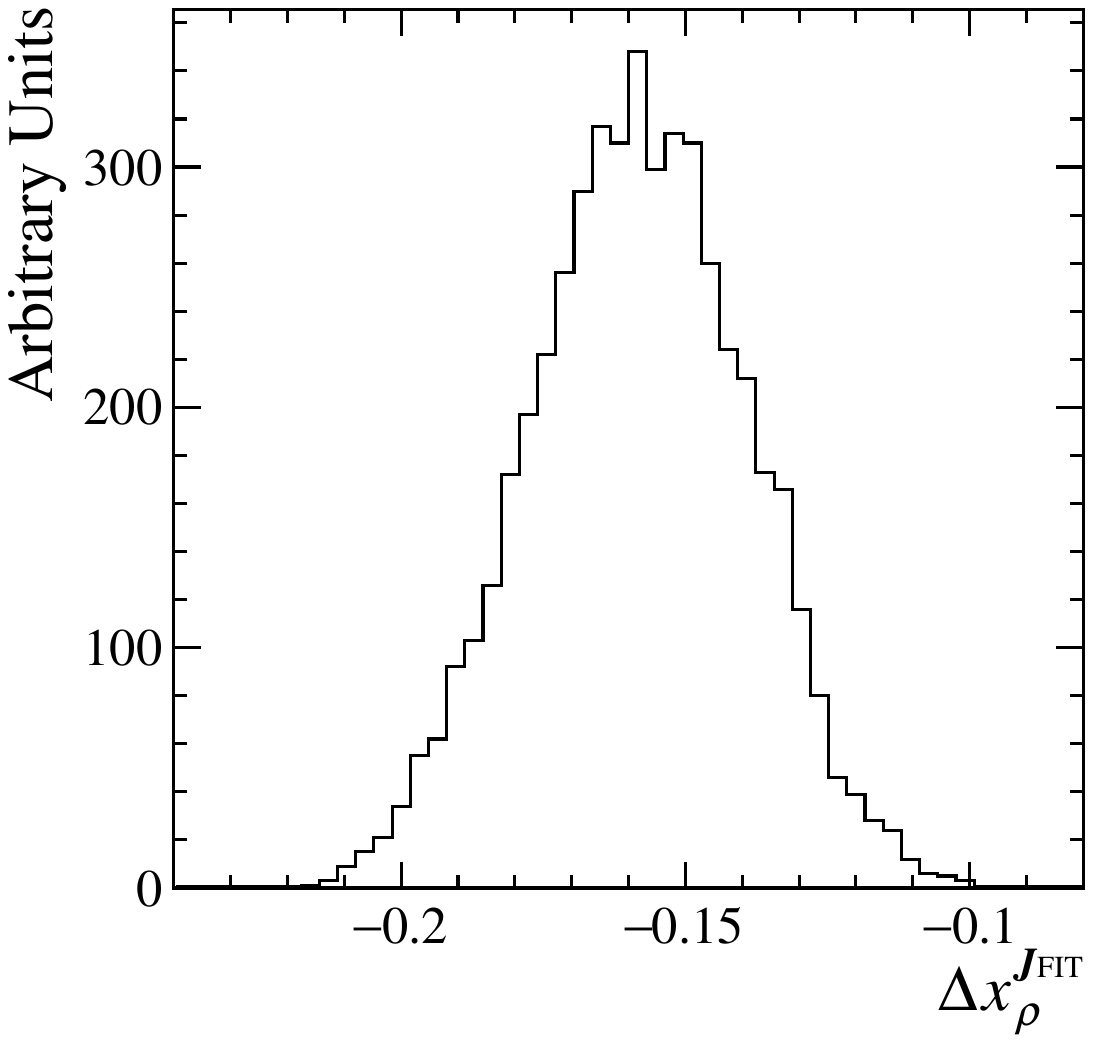}
\caption{\small
Distributions of $\dxrho$ results obtained by (top left) na\"ive averages taking into account the asymmetric uncertainties of the individual fits to two samples, (bottom right) joint fits performed in the \jfit framework, (top right) the former versus the latter, and (bottom left) the difference between the two. 
\label{fig:compDeltaX}}
\end{figure}

The results obtained from methods 1 and 2 above are found to be equivalent: in the present case, with a very few exceptions, the effect of the asymmetric nature of the likelihood is negligible comparing to the statistical uncertainty.
The results obtained from methods 2 and 3 are compared in Fig.~\ref{fig:compDeltaX}, and show a much larger difference.
The distribution of the difference between results obtained by the two methods has a full width at half maximum of approximately $25\%$ of the typical statistical uncertainty on $\dxrho$ in fits to individual datasets (which is $0.03$).

\begin{figure}[htbp]
\includegraphics[width=0.5\textwidth]{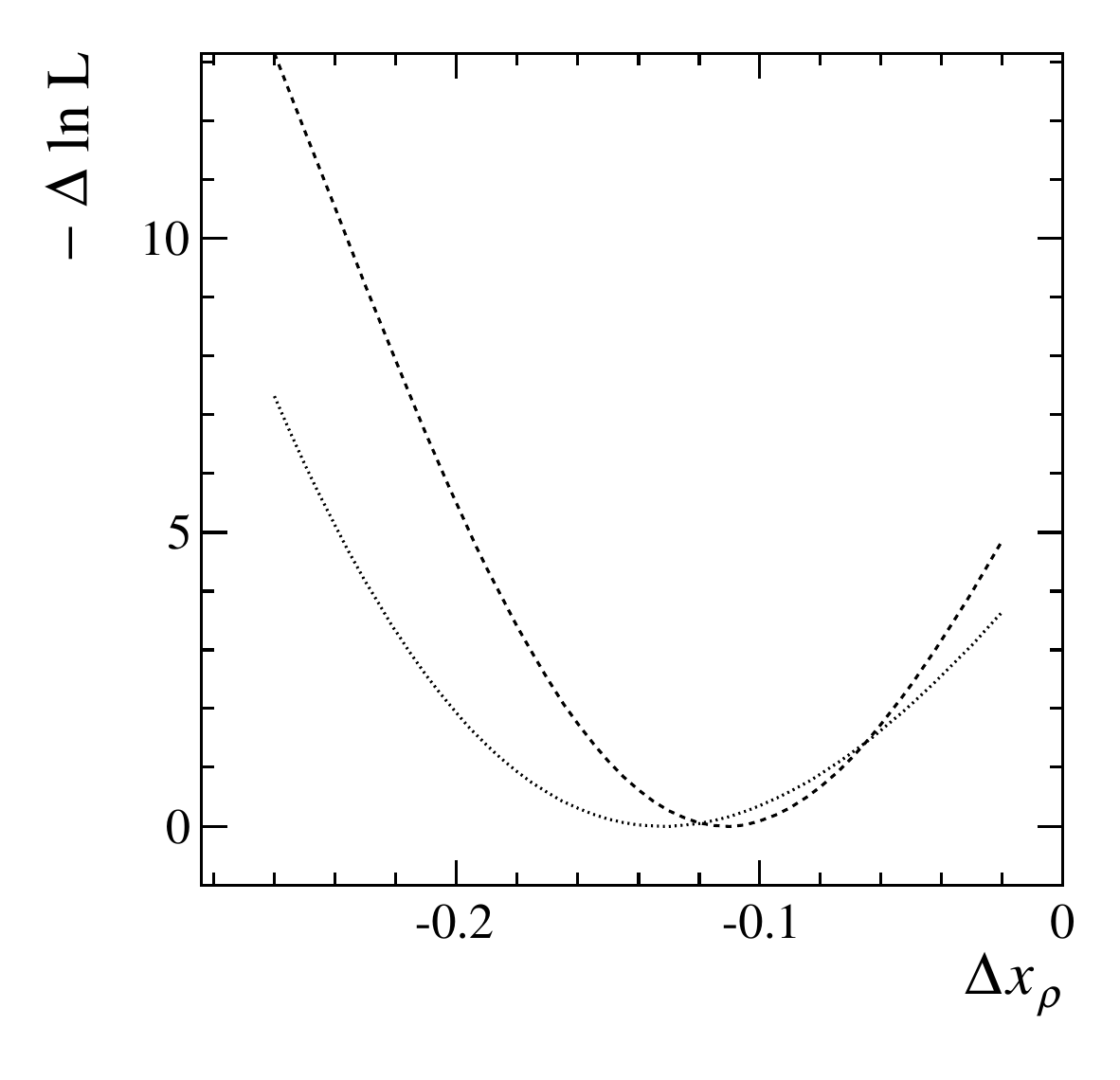}
\includegraphics[width=0.5\textwidth]{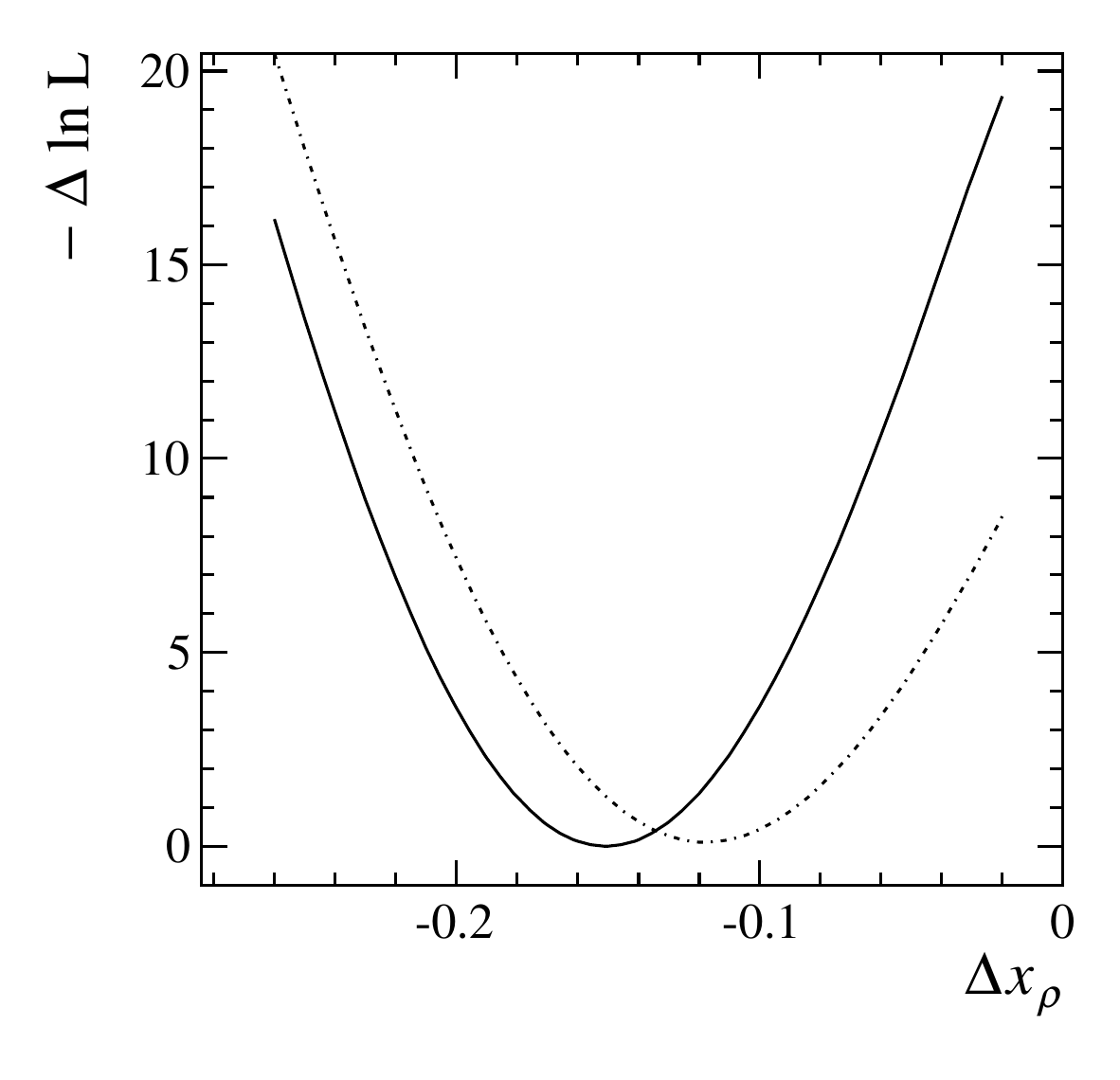}
\caption{\small
Left: log-likelihood scans, showing $-\Delta \ln \L \equiv (\ln \L)_{\rm min}- \ln \L$ as a function of $\dxrho$ in two different datasets.
Right: the sum of these two log-likelihoods scans (dashed-dotted curve), compared to the scan obtained from a joint fit to the two samples, performed in the \jfit framework (solid curve).
It should be noted that the result obtained from the joint fit is more negative than both of those from the individual fits and is closer to the true value ($-0.16$).
In addition it has a smaller uncertainty than the simple average.
\label{fig:scans}}
\end{figure}

We perform likelihood scans as a function of $\dxrho$ for several individual datasets and their corresponding joint fits, i.e., we fix $\dxrho$ to several consecutive values, for each of which the fit is repeated.
The other parameters are free to vary as in the nominal fit.
We compare the sum of one-dimensional log-likelihood functions obtained from scans of two datasets, to the full likelihood scan obtained with the corresponding \jfit{}-framework fit. 
One such comparison, which is more extreme than the average case, but not uncommon, is shown in Fig.~\ref{fig:scans}.
It illustrates the fact that even if two experiments provide their likelihood dependences on a particular subset of parameters, obtaining a combined result by summing these functions is in general not equivalent to performing a joint fit.
Indeed, values of nuisance parameters, for which combined results are not desired, generally differ between the joint fit and the individual fits.
Note that in the particular case shown in Fig.~\ref{fig:scans} the minimum obtained from the joint fit does not lie between the two minima obtained from the individual fits but is located at a more negative value and, in fact, is closer to the generated value of $-0.16$. It also has a smaller uncertainty than the other combination.
This indicates that even in well-behaved cases and even if the combination is performed by summing partial likelihood functions, neglecting correlations may result in biases.

\begin{figure}[htbp]
\includegraphics[width=0.5\textwidth]{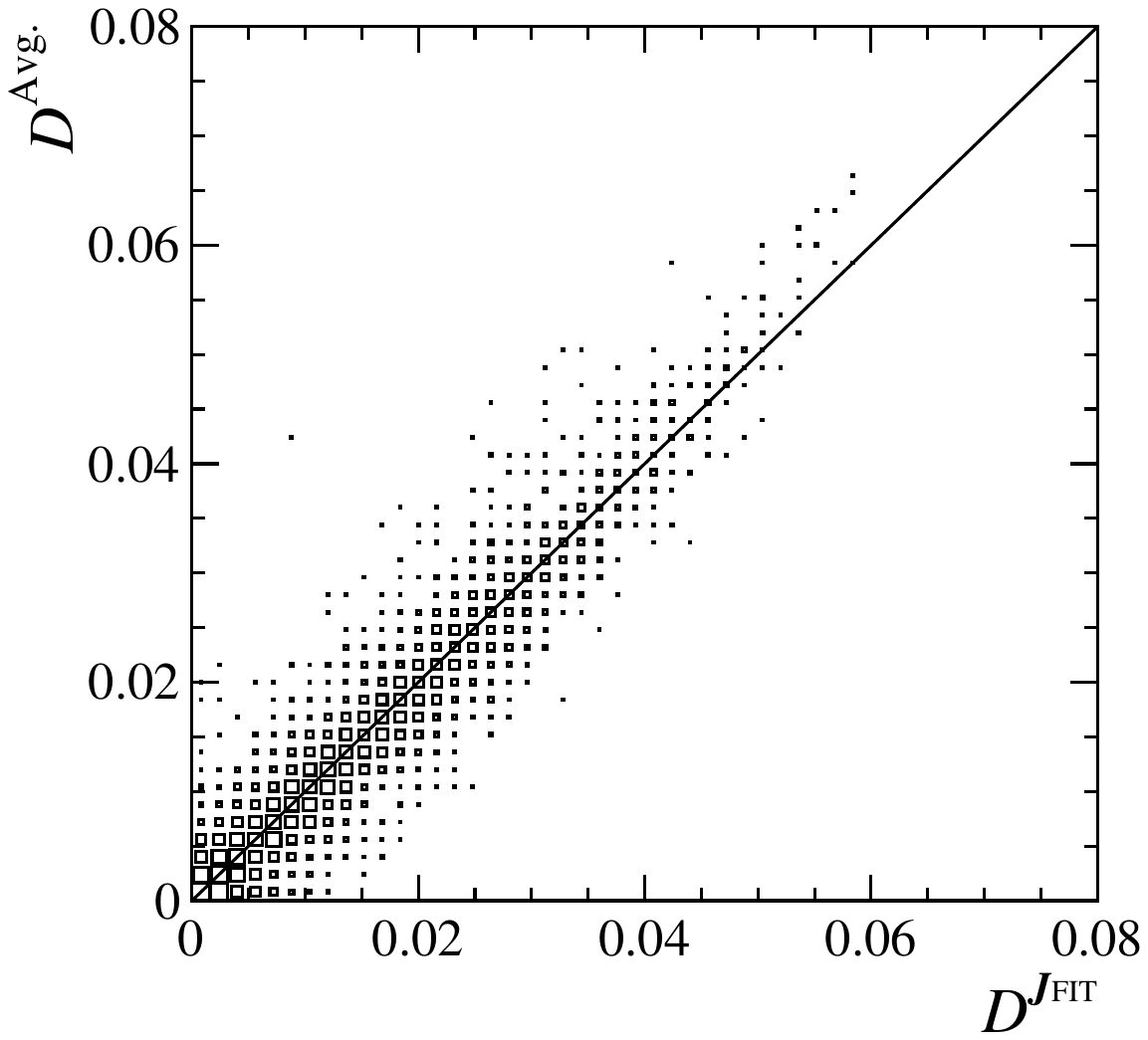}
\includegraphics[width=0.5\textwidth]{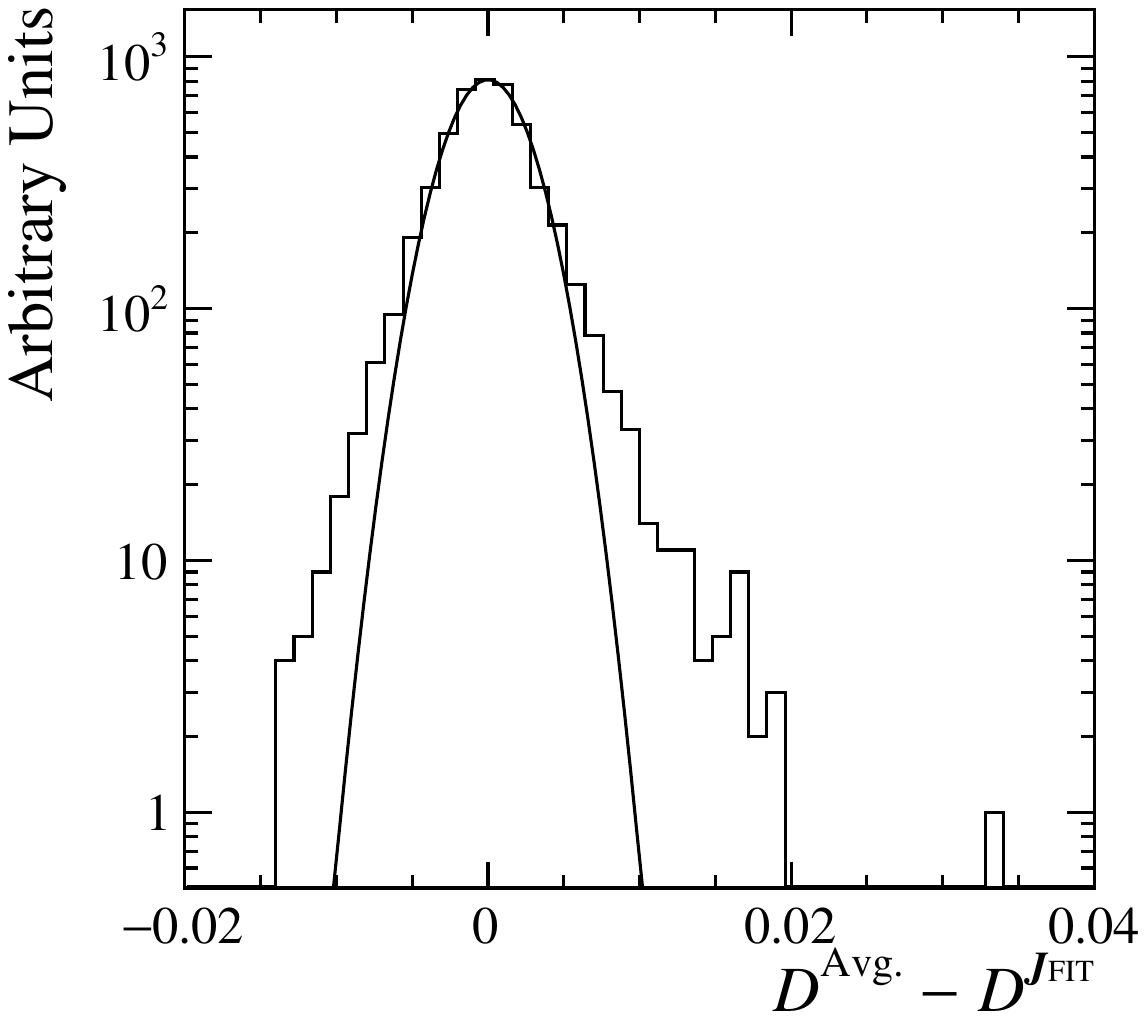}
\caption{\small
Left: distance from the generated value ($-0.16$) of $\dxrho$ results obtained by na\"ive averaging versus that corresponding to \jfit{}-framework fits.
Right: distribution of the difference between the former and the latter. The solid smooth curve is a Gaussian fitted to the central region of the distribution $[-0.005, 0.005]$, to clarify the presence of the non-Gaussian tails.
\label{fig:distFromTrue}}
\end{figure}

\begin{figure}[htbp]
\centering
\includegraphics[width=0.55\textwidth]{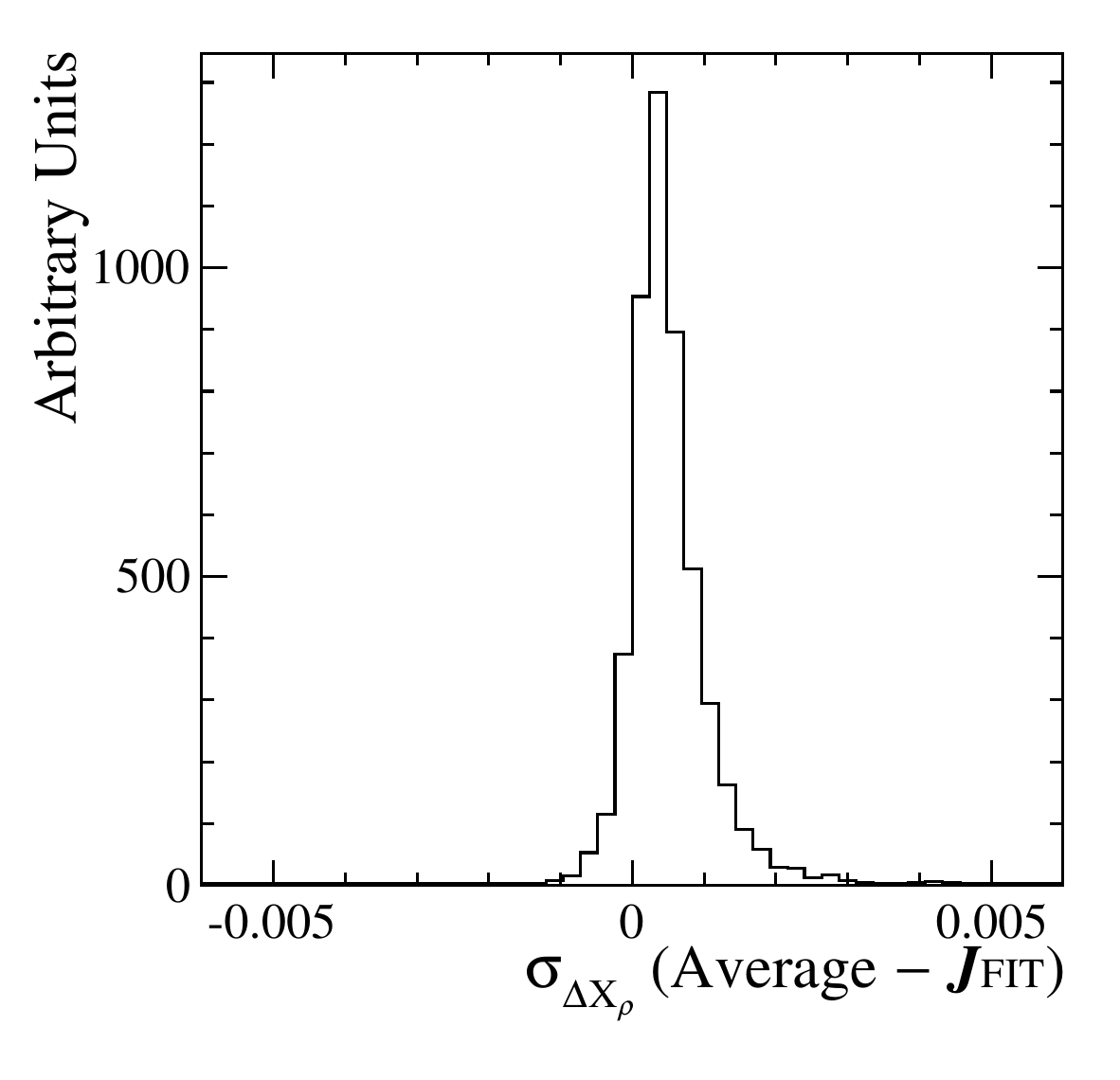}
\caption{\small
The difference between uncertainties obtained in na\"ive averages, and those from joint fits performed in the \jfit framework.
In the former, the average between the positive and negative asymmetric uncertainties is used. In $88\%$ of the cases joint fits yield improved sensitivity to $\dxrho$.
\label{fig:compErrors}}
\end{figure}

To evaluate how often one of the features illustrated in Fig.~\ref{fig:scans} occurs,
the distance of combined results to the generated value of $\dxrho$ has been studied.
For each of the 4950 pairwise combinations of datasets, we compute $D^{\rm Avg.} = \left| \dxrhoavg - \dxrhogen \right|$, and $D^{\sjfit} = \left| \dxrhojfit - \dxrhogen \right|$,
where $\dxrhogen=-0.16$ is the generated value of $\dxrho$. The distances $D^{\rm Avg.}$ and $D^{\sjfit}$ are then compared.
This study shows that results obtained by joint fits are more often closer to the generated value than those obtained by na\"ive averages due to the fact that they fully account for the correlations between the fit variables.
Figure~\ref{fig:distFromTrue} shows $D^{\rm Avg.}$ versus $D^{\sjfit}$ in the different pairwise combinations and the distribution of the difference between the former and the latter.
This example shows that, while both methods can yield unbiased results, joint fits are more often closer to the true value.
Moreover, to clarify the presence of the non-Gaussian tails in the distribution of differences, it is overlaid with a Gaussian fitted to its central region $[-0.005, 0.005]$; numbers of positive and negative entries in the distribution, excluding the ranges corresponding to one, two and three standard deviations of the Gaussian are given in Table~\ref{tab:posNegEntries}.
Comparison of the statistical uncertainties obtained from na\"ive averages and joint fits are shown in Fig.~\ref{fig:compErrors}.
In $88\%$ of the cases the uncertainty obtained from a joint fit is smaller.

\begin{table}[htbp]
\caption{\small
The distribution on the right hand side of Fig.~\ref{fig:distFromTrue} illustrates the fact that results obtained by joint fits are more often closer to the generated value than these obtained by na\"ive averages.
Here are given the numbers of positive and negative entries in the distribution, excluding the ranges corresponding to one, two and three standard deviations ($\sigma$) of the overlaid Gaussian.
\label{tab:posNegEntries}}
\centering
\begin{tabular}{|c|c|c|}
  \hline
  Excluded          & Number of        & Number of        \\
  region ($\sigma$) & positive entries & negative entries \\
  \hline
  3 & 126 & 71 \\
  2 & 330 & 266 \\
  1 & 918 & 911 \\
  \hline
\end{tabular}

\end{table}

We stress that the example given here is not extreme: likelihoods are nearly Gaussian and are rather well behaved, as can be seen in Fig.~\ref{fig:scans}.
In cases where the likelihood presents strong non-linear features, such as asymmetric functions that cannot be well described by a bifurcated Gaussian, or if it has multiple minima, the difference between na\"ive averaging and joint fitting could be much larger.
In practice, multiple solutions appear in nearly all the Dalitz-plot analyses performed by the \B factories; they represent one of the major difficulties in these analyses.
Clearly, a joint fit allows to resolve better the global minimum from the mirror solutions.

\clearpage

\section{Example worker class}
\label{sec:roofit-slave}

In this section we give an example implementation of a worker class, \texttt{LauRooFitSlave}, that is derived from the \texttt{LauSimFitSlave} base class.
The particular implementation uses classes from the \roofit framework~\cite{ref:roofit} to describe the fit model, store the data to be fitted and to evaluate the likelihood function.
It is sufficiently general to cover the majority of \roofit-based fitting scenarios and can be quite straightforwardly extended to include those with conditional observables, fitting only subsets of the data, etc.
The implementations of the constructor and destructor, each of the eight pure virtual member functions mentioned in Section~\ref{sec:jfitFramework}, as well as some additional utility functions, are given with some accompanying explanatory text.
For the full documentation and source code please see Ref.~\cite{ref:rf-slave}.

\subsection{Class data members}

The data members of the class are as follows:
\begin{lstlisting}
        //! The fit model
        RooAbsPdf& model_;

        //! The dataset variables
        RooArgSet dataVars_;

        //! The name of the (optional) weight variable in the dataset
        TString weightVarName_;

        //! The data file
        TFile* dataFile_;

        //! The data tree
        TTree* dataTree_;

        //! The data for the current experiment
        RooAbsData* exptData_;

        //! Is the PDF extended?
        const Bool_t extended_;

        //! The experiment category variable
        RooCategory iExptCat_;

        //! The NLL variable
        RooNLLVar* nllVar_;

        //! The fit parameters (as RooRealVar's)
        std::vector<RooRealVar*> fitVars_;

        //! The fit parameters (as LauParameter's)
        std::vector<LauParameter*> fitPars_;
\end{lstlisting}

\subsection{Constructor and destructor}

The constructor takes as arguments the fit model, a flag to indicate whether or not the fit is an extended fit, the fit variables, and the name of the variable in the data that should be used as an event-by-event weight (if any):
\begin{lstlisting}
LauRooFitSlave::LauRooFitSlave( RooAbsPdf& model,
                                const Bool_t extended,
                                const RooArgSet& vars,
                                const TString& weightVarName ) :
        LauSimFitSlave(),
        model_(model),
        dataVars_(vars),
        weightVarName_(weightVarName),
        dataFile_(0),
        dataTree_(0),
        exptData_(0),
        extended_(extended),
        iExptCat_("iExpt","Expt Number"),
        nllVar_(0)
{
}
\end{lstlisting}
The destructor cleans up any allocated memory:
\begin{lstlisting}
LauRooFitSlave::~LauRooFitSlave()
{
        delete nllVar_; nllVar_ = 0;
        this->cleanData();
}
\end{lstlisting}

\subsection{Utility functions}

The \texttt{cleanData} utility function cleans up the memory associated with the data storage.
The \texttt{convertToLauParmaeter} and \texttt{convertToLauParmaeters} functions convert the \roofit versions of the fit parameters (either \texttt{RooRealVar} or \texttt{RooFormulaVar} objects) into \texttt{LauParameter} objects.

\begin{lstlisting}
void LauRooFitSlave::cleanData()
{
        if ( dataFile_ != 0 ) {
                dataFile_->Close();
                delete dataFile_;
                dataTree_ = 0;
                dataFile_ = 0;
        }
        delete exptData_;
        exptData_ = 0;
}
\end{lstlisting}

\begin{lstlisting}
LauParameter* LauRooFitSlave::convertToLauParameter( const RooRealVar* rooParameter ) const
{
        return new LauParameter( rooParameter->GetName(), rooParameter->getVal(), rooParameter->getMin(), rooParameter->getMax(), rooParameter->isConstant() );
}
\end{lstlisting}

\begin{lstlisting}
std::vector< std::pair<RooRealVar*,LauParameter*> > LauRooFitSlave::convertToLauParameters( const RooFormulaVar* rooFormula ) const
{
        // Create the empty vector
        std::vector< std::pair<RooRealVar*,LauParameter*> > lauParameters;

        Int_t parIndex(0);
        RooAbsArg* rabsarg(0);
        RooRealVar* rrvar(0);
        RooFormulaVar* rfvar(0);
        // Loop through all the parameters of the formula
        while ( (rabsarg = rooFormula->getParameter(parIndex)) ) {
                // First try converting to a RooRealVar
                rrvar = dynamic_cast<RooRealVar*>( rabsarg );
                if ( rrvar ) {
                        // Do the conversion and add it to the array
                        LauParameter* lpar = this->convertToLauParameter( rrvar );
                        lauParameters.push_back( std::make_pair(rrvar,lpar) );
                        continue;
                }

                // If that didn't work, try converting to a RooFormulaVar
                rfvar = dynamic_cast<RooFormulaVar*>( rabsarg );
                if ( rfvar ) {
                        // Do the conversion and add these to the array
                        std::vector< std::pair<RooRealVar*,LauParameter*> > lpars = this->convertToLauParameters( rfvar );
                        for ( std::vector< std::pair<RooRealVar*,LauParameter*> >::iterator iter = lpars.begin(); iter != lpars.end(); ++iter ) {
                                lauParameters.push_back( *iter );
                        }
                        continue;
                }

                // If neither of those worked we don't know what to do, so print an error message and continue
                std::cerr << "ERROR in LauRooFitSlave::convertToLauParameters : One of the parameters is not a RooRealVar nor a RooFormulaVar, it is a: " << rabsarg->ClassName() << std::endl;
                std::cerr << "                                                : Do not know how to process that - it will be skipped." << std::endl;
        }

        return lauParameters;
}
\end{lstlisting}

\subsection{The \texttt{initialise} function}

\begin{lstlisting}
void LauRooFitSlave::initialise()
{
        if ( weightVarName_ != "" ) {
                Bool_t weightVarFound = kFALSE;
                RooFIter argset_iter = dataVars_.fwdIterator();
                RooAbsArg* param(0);
                while ( (param = argset_iter.next()) ) {
                        TString name = param->GetName();
                        if ( name == weightVarName_ ) {
                                weightVarFound = kTRUE;
                                break;
                        }
                }
                if ( ! weightVarFound ) {
                        std::cerr << "ERROR in LauRooFitSlave::initialise : The set of data variables does not contain the weighting variable \"" << weightVarName_ << std::endl;
                        std::cerr << "                                    : Weighting will be disabled." << std::endl;
                        weightVarName_ = "";
                }
        }
}
\end{lstlisting}

\subsection{The \texttt{verifyFitData} function}

\begin{lstlisting}
Bool_t LauRooFitSlave::verifyFitData(const TString& dataFileName, const TString& dataTreeName)
{
        // Clean-up from any previous runs
        if ( dataFile_ != 0 ) {
                this->cleanData();
        }

        // Open the data file
        dataFile_ = TFile::Open( dataFileName );
        if ( ! dataFile_ ) {
                std::cerr << "ERROR in LauRooFitSlave::verifyFitData : Problem opening data file \"" << dataFileName << "\"" << std::endl;
                return kFALSE;
        }

        // Retrieve the tree
        dataTree_ = dynamic_cast<TTree*>( dataFile_->Get( dataTreeName ) );
        if ( ! dataTree_ ) {
                std::cerr << "ERROR in LauRooFitSlave::verifyFitData : Problem retrieving tree \"" << dataTreeName << "\" from data file \"" << dataFileName << "\"" << std::endl;
                dataFile_->Close();
                delete dataFile_;
                dataFile_ = 0;
                return kFALSE;
        }

        // Check that the tree contains branches for all the fit variables
        RooFIter argset_iter = dataVars_.fwdIterator();
        RooAbsArg* param(0);
        Bool_t allOK(kTRUE);
        while ( (param = argset_iter.next()) ) {
                TString name = param->GetName();
                TBranch* branch = dataTree_->GetBranch( name );
                if ( branch == 0 ) {
                        std::cerr << "ERROR in LauRooFitSlave::verifyFitData : The data tree does not contain a branch for fit variable \"" << name << std::endl;
                        allOK = kFALSE;
                }
        }
        if ( ! allOK ) {
                return kFALSE;
        }

        // Check whether the tree has the branch iExpt
        TBranch* branch = dataTree_->GetBranch("iExpt");
        if ( branch == 0 ) {
                std::cout << "WARNING in LauRooFitSlave::verifyFitData : Cannot find branch \"iExpt\" in the tree, will treat all data as being from a single experiment" << std::endl;
        } else {
                // Define the valid values for the iExpt RooCategory
                iExptCat_.clearTypes();
                const UInt_t firstExp = dataTree_->GetMinimum("iExpt");
                const UInt_t lastExp  = dataTree_->GetMaximum("iExpt");
                for ( UInt_t iExp = firstExp; iExp <= lastExp; ++iExp ) {
                        iExptCat_.defineType( TString::Format("expt%d",iExp), iExp );
                }
        }

        return kTRUE;
}
\end{lstlisting}

\subsection{The \texttt{prepareInitialParArray} function}

\begin{lstlisting}
void LauRooFitSlave::prepareInitialParArray( TObjArray& array )
{
        // Check that the NLL variable has been initialised
        if ( ! nllVar_ ) {
                std::cerr << "ERROR in LauRooFitSlave::prepareInitialParArray : NLL var not initialised" << std::endl;
                return;
        }

        // If we already prepared the entries in the fitPars_ vector then we only need to add the contents to the array
        if ( ! fitPars_.empty() ) {
                for ( std::vector<LauParameter*>::iterator iter = fitPars_.begin(); iter != fitPars_.end(); ++iter ) {
                        array.Add(*iter);
                }
                return;
        }

        // Store the set of parameters and the total number of parameters
        RooArgSet* varSet = nllVar_->getParameters( exptData_ );
        UInt_t nFreePars(0);

        // Loop through the fit parameters
        RooFIter argset_iter = varSet->fwdIterator();
        RooAbsArg* param(0);
        while ( (param = argset_iter.next()) ) {
                // Only consider the free parameters
                if ( ! param->isConstant() ) {
                        // Add the parameter
                        RooRealVar* rrvar = dynamic_cast<RooRealVar*>( param );
                        if ( rrvar != 0 ) {
                                // Count the number of free parameters
                                ++nFreePars;
                                // Do the conversion and add it to the array
                                LauParameter* lpar = this->convertToLauParameter( rrvar );
                                fitVars_.push_back( rrvar );
                                fitPars_.push_back( lpar );
                                array.Add( lpar );
                        } else {
                                RooFormulaVar* rfvar = dynamic_cast<RooFormulaVar*>( param );
                                if ( rfvar == 0 ) {
                                        std::cerr << "ERROR in LauRooFitSlave::prepareInitialParArray : The parameter is neither a RooRealVar nor a RooFormulaVar, don't know what to do" << std::endl;
                                        continue;
                                }
                                std::vector< std::pair<RooRealVar*,LauParameter*> > lpars = this->convertToLauParameters( rfvar );
                                for ( std::vector< std::pair<RooRealVar*,LauParameter*> >::iterator iter = lpars.begin(); iter != lpars.end(); ++iter ) {
                                        RooRealVar* rrv = iter->first;
                                        LauParameter* lpar = iter->second;
                                        if ( ! rrv->isConstant() ) {
                                                continue;
                                        }

                                        // Count the number of free parameters
                                        ++nFreePars;
                                        // Add the parameter to the array
                                        fitVars_.push_back( rrvar );
                                        fitPars_.push_back( lpar );
                                        array.Add( lpar );
                                }
                        }
                }
        }
        delete varSet;

        this->startNewFit( nFreePars, nFreePars );
}
\end{lstlisting}

\subsection{The \texttt{getTotNegLogLikelihood} function}

\begin{lstlisting}
Double_t LauRooFitSlave::getTotNegLogLikelihood()
{
        Double_t nLL = (nllVar_ != 0) ? nllVar_->getVal() : 0.0;
        return nLL;
}
\end{lstlisting}

\subsection{The \texttt{setParsFromMinuit} function}

\begin{lstlisting}
void LauRooFitSlave::setParsFromMinuit(Double_t* par, Int_t npar)
{
        // This function sets the internal parameters based on the values
        // that Minuit is using when trying to minimise the total likelihood function.

        // MINOS reports different numbers of free parameters depending on the
        // situation, so disable this check
        const UInt_t nFreePars = this->nFreeParams();
        if ( ! this->withinAsymErrorCalc() ) {
                if (static_cast<UInt_t>(npar) != nFreePars) {
                        std::cerr << "ERROR in LauRooFitSlave::setParsFromMinuit : Unexpected number of free parameters: " << npar << ".\n";
                        std::cerr << "                                             Expected: " << nFreePars << ".\n" << std::endl;
                        gSystem->Exit(EXIT_FAILURE);
                }
        }

        // Despite npar being the number of free parameters
        // the par array actually contains all the parameters,
        // free and floating...

        // Update all the floating ones with their new values
        for (UInt_t i(0); i<nFreePars; ++i) {
                if (!fitPars_[i]->fixed()) {
                        // Set both the RooRealVars and the LauParameters
                        fitPars_[i]->value(par[i]);
                        fitVars_[i]->setVal(par[i]);
                }
        }
}
\end{lstlisting}

\subsection{The \texttt{readExperimentData} function}

\begin{lstlisting}
UInt_t LauRooFitSlave::readExperimentData()
{
        // check that we're being asked to read a valid index
        const UInt_t exptIndex = this->iExpt();
        if ( iExptCat_.numTypes() == 0 && exptIndex != 0 ) {
                std::cerr << "ERROR in LauRooFitSlave::readExperimentData : Invalid experiment number " << exptIndex << ", data contains only one experiment" << std::endl;
                return 0;
        } else if ( ! iExptCat_.isValidIndex( exptIndex ) ) {
                std::cerr << "ERROR in LauRooFitSlave::readExperimentData : Invalid experiment number " << exptIndex << std::endl;
                return 0;
        }

        // cleanup the data from any previous experiment
        delete exptData_;

        // retrieve the data and find out how many events have been read
        if ( iExptCat_.numTypes() == 0 ) {
                exptData_ = new RooDataSet( TString::Format("expt%dData",exptIndex), "", dataTree_, dataVars_, "", (weightVarName_ != "") ? weightVarName_.Data() : 0 );
        } else {
                const TString selectionString = TString::Format("iExpt==%d",exptIndex);
                TTree* exptTree = dataTree_->CopyTree(selectionString);
                exptData_ = new RooDataSet( TString::Format("expt%dData",exptIndex), "", exptTree, dataVars_, "", (weightVarName_ != "") ? weightVarName_.Data() : 0 );
                delete exptTree;
        }

        const UInt_t nEvent = exptData_->numEntries();
        this->eventsPerExpt( nEvent );
        return nEvent;
}
\end{lstlisting}

\begin{lstlisting}
void LauRooFitSlave::cacheInputFitVars()
{
        // cleanup the old NLL info
        delete nllVar_;

        // construct the new NLL variable for this dataset
        nllVar_ = new RooNLLVar("nllVar", "", model_, *exptData_, extended_);
}
\end{lstlisting}

\subsection{The \texttt{finaliseExperiment} function}

\begin{lstlisting}
void LauRooFitSlave::finaliseExperiment( const LauAbsFitter::FitStatus& fitStat, const TObjArray* parsFromMaster, const TMatrixD* covMat, TObjArray& parsToMaster )
{
        // Copy the fit status information
        this->storeFitStatus( fitStat, *covMat );

        // Now process the parameters
        const UInt_t nFreePars = this->nFreeParams();
        UInt_t nPars = parsFromMaster->GetEntries();
        if ( nPars != nFreePars ) {
                std::cerr << "ERROR in LauRooFitSlave::finaliseExperiment : Unexpected number of parameters received from master" << std::endl;
                std::cerr << "                                            : Received " << nPars << " when expecting " << nFreePars << std::endl;
                gSystem->Exit( EXIT_FAILURE );
        }

        for ( UInt_t iPar(0); iPar < nPars; ++iPar ) {
                LauParameter* parameter = dynamic_cast<LauParameter*>( (*parsFromMaster)[iPar] );
                if ( ! parameter ) {
                        std::cerr << "ERROR in LauRooFitSlave::finaliseExperiment : Error reading parameter from master" << std::endl;
                        gSystem->Exit( EXIT_FAILURE );
                }

                if ( parameter->name() != fitPars_[iPar]->name() ) {
                        std::cerr << "ERROR in LauRooFitSlave::finaliseExperiment : Error reading parameter from master" << std::endl;
                        gSystem->Exit( EXIT_FAILURE );
                }

                *(fitPars_[iPar]) = *parameter;

                RooRealVar* rrv = fitVars_[iPar];
                rrv->setVal( parameter->value() );
                rrv->setError( parameter->error() );
                rrv->setAsymError( parameter->negError(), parameter->posError() );
        }

        // Update the pulls and add each finalised fit parameter to the list to
        // send back to the master
        for ( std::vector<LauParameter*>::iterator iter = fitPars_.begin(); iter != fitPars_.end(); ++iter ) {
                (*iter)->updatePull();
                parsToMaster.Add( *iter );
        }

        // Write the results into the ntuple
        std::vector<LauParameter> extraVars;
        LauFitNtuple* ntuple = this->fitNtuple();
        ntuple->storeParsAndErrors(fitPars_, extraVars);

        // find out the correlation matrix for the parameters
        ntuple->storeCorrMatrix(this->iExpt(), this->fitStatus(), this->covarianceMatrix());

        // Fill the data into ntuple
        ntuple->updateFitNtuple();
}
\end{lstlisting}

\end{document}